# Thermochemical non-equilibrium effects on turbulent boundary layers

**Jun-Yang Li[1], Ming Yu[1], Dong Sun[1], Hong-Min Su[1], Peng-Xin Liu[1], Xian-Xu Yuan[1]**

[1]State Key Laboratory of Aeradynamics, Mianyang 621000, China

This investigation employs direct numerical simulations (DNS) of high-Mach-number turbulent boundary layers under three flow conditions: a low-enthalpy calorically perfect gas, and two high-temperature gas mixtures, one in chemical non-equilibrium state and the other in full thermochemical non-equilibrium state. The influences of the two-temperature model on turbulent statistics and the coupling among turbulence, chemistry, and vibrational energy are examined. It is found that while high-enthalpy effects leave the velocity statistics virtually unchanged, they dramatically modify the near-wall temperature field. A pronounced disparity between the translational-rotational temperature and the vibrational temperature arises in the near-wall region, rendering the conventional generalized Reynolds analogy (GRA) inaccurate for vibrational temperature. To remedy this, a novel composite GRA is proposed that blends a vibrational-temperature-based relation with the standard formulation, and it demonstrates excellent agreement with the DNS data. Thermal non-equilibrium effects also substantially alter near-wall chemical reactions: it suppresses $O_2$ dissociation while promoting NO formation, leading to a corresponding decrease and increase in the mean concentrations of O and NO, respectively. Spectral analyses of the turbulence-chemistry and turbulence-vibrational relaxation interaction terms reveal that temperature fluctuations dominate these flow quantities at energy-containing scales. Integrating the resulting spectral functions, we evaluate subgrid-scale closure terms for large-eddy simulation. At small filtering scales, the magnitude of the cross-correlation term rivals or exceeds that of the temperature fluctuation term, whereas the temperature fluctuation term becomes dominant at larger filter scales. These findings deliver fundamental physical insights and robust closures indispensable for the predictive modeling of high-enthalpy compressible turbulent flows.

## 1. Introduction

Compressible turbulent boundary layers (TBLs) are ubiquitous over the fuselages of high-speed aerospace vehicles. Their statistical properties and underlying transport mechanisms dictate the fidelity of aerodynamic drag and aerothermal heating predictions(Anderson 2006). Consequently, the physics of these flows has remained a focal point of research, especially in high-Mach-number regimes(Duan *et al.* 2010; Xu *et al.* 2021; Huang *et al.* 2022; Cogo *et al.* 2023). The maturation of direct numerical simulations (DNS) has enabled the establishment of extensive high-fidelity databases, significantly advancing our comprehension of the flow mechanisms and facilitating the development of advanced turbulence models (Pirozzoli & Bernardini 2011; Duan *et al.* 2011; Zhang *et al.* 2018; Huang *et al.* 2020; Cogo *et al.* 2022; Huang *et al.* 2022; Gibis *et al.* 2024; Szajnecki *et al.* 2025).

Substantial progress has been achieved in characterizing compressible TBLs. A primary aspect of research is the development of scaling laws, or compressible transformations, aimed at collapsing the statistical behavior of compressible flows onto their incompressible counterparts. In recent years, the formulation of integral transformations for the mean velocity profile has garnered significant attention (van Driest 1951; Trettel & Larsson 2016; Volpiani *et al.* 2020; Griffin *et al.* 2021; Hasan *et al.* 2023). These transformations successfully map the compressible mean velocity to the classic linear law in the viscous sublayer and the logarithmic law by accounting for mean density and dynamic viscosity variations. For Reynolds stresses, semi-local scalings (Huang *et al.* 1995; Foysi *et al.* 2004) have been introduced to mitigate the sensitivity of their profiles to the Mach number and wall thermal conditions (Lagha *et al.* 2011; Duan *et al.* 2011; Zhang *et al.* 2018). Furthermore, the thermal and momentum fields are statistically coupled based on the resemblance between the mean momentum and total enthalpy equations. This coupling is captured by the strong Reynolds analogy (Gaviglio 1987; Huang *et al.* 1995), which was subsequently refined into the generalized Reynolds analogy (GRA) by Zhang *et al.* (2014) by substituting the conventional recovery temperature with a generalized recovery temperature. Comprehensive reviews of these analytical frameworks have been compiled recently by Cheng & Fu (2025) and Yu *et al.* (2024), and will not be detailed further herein.

As the freestream Mach number increases further, intense aerodynamic heating within the boundary layer elevates the static temperature to levels at which high-enthalpy effects, such as vibrational energy excitation, and molecular dissociation of oxygen and nitrogen, become non-negligible. The flow thus becomes a multi-component, chemically reacting gas mixture (Anderson 2006). Because the characteristic time scales of turbulent mixing, vibrational relaxation, and chemical reactions are comparable, the boundary layer is in a state of pronounced thermochemical non-equilibrium, distinguished by a two-way coupling between turbulence and finite-rate physicochemical processes (Urzay & Renzo 2021). Such flows are termed high-enthalpy turbulent boundary layers (TBLs), in contrast to the traditional calorically perfect gas, which relies on simplifying thermal and chemical equilibrium assumptions and neglects non-equilibrium phenomena (Duan & Martin 2011*b*).

To resolve these physics, Duan & Martin (2011*b*) performed the pioneering DNS of a high-enthalpy turbulent boundary layer, revealing that mean velocity statistics are remarkably insensitive to chemical non-equilibrium, whereas the mean temperature and density profiles are substantially modified by endothermic reactions. They further established that Morkovin's hypothesis, the van Driest transformation, and the strong Reynolds analogy (SRA) preserve their validity under high-enthalpy conditions. Subsequently, Renzo & Urzay (2021) investigated the laminar-to-turbulent transition and the fully developed regime of a Mach 10, chemically non-equilibrium boundary layer. Passiatore *et al.* (2021) compared chemically non-equilibrium and chemically frozen conditions, concluding that chemical reactions modify the mean thermodynamic field predominantly through endothermic forward pathways while exerting negligible influence on the first- and second-order velocity statistics.

Further investigations by Liu *et al.* (2022, 2023) and Li *et al.* (2022) have elucidated the turbulent transport mechanisms of individual chemical species and their contributions to the wall heat flux. To quantify the coupling between fluid dynamics and chemical reactions, Duan & Martín (2011) and Duan & Martin (2011*a*) introduced two governing parameters: the interaction Damköhler number and the interaction heat release. Within this framework, Renzo & Urzay (2021) identified atomic and molecular oxygen O and $O_2$ as the primary species participating in these interactions, while Williams *et al.* (2025) demonstrated that temperature fluctuations, species density fluctuations, and their joint statistics contribute significantly to the net chemical production rates. To model these effects, Wang & Xu (2026) recognized temperature fluctuations as the dominant component of their interactions and developed a probability density function (PDF) closure model for the averaged chemical source term.

Under conditions of thermochemical non-equilibrium, the vibrational temperature emerges as an independent thermodynamic state variable, distinct from the translational-rotational temperature. To the best of the authors' knowledge, Passiatore *et al.* (2022) conducted the only DNS to date that explicitly resolves thermal non-equilibrium via a two-temperature framework, specifically for a Mach 12 turbulent boundary layer over a non-catalytic cold wall. Their findings revealed substantial local thermodynamic splitting (deviations) between the translational-rotational and vibrational temperatures in the near-wall region, followed by a gradual relaxation toward thermal equilibrium in the outer layer. Furthermore, by introducing a vibrational turbulent Prandtl number, they demonstrated that the strong Reynolds analogy remains approximately valid, while noting that the $N_2$-dominated vibrational relaxation is governed by a relatively long characteristic timescale.

This brief literature review highlights that a quantitative assessment of the individual and combined effects of thermal and chemical non-equilibrium on turbulent boundary layer statistics is still lacking, which obscures their relative contributions to wall heat fluxes and the underlying flow physics. Furthermore, it remains an open question whether the vibrational temperature obeys a scaling law analogous to the generalized Reynolds analogy established for the translational-rotational temperature. Simultaneously, the influence of thermal non-equilibrium on turbulence-chemistry interactions, as well as the fundamental nature of turbulence-vibrational relaxation interactions, has yet to be systematically investigated. This serves as the motivation for the present study. In this work, we aim to address these open questions by performing comparative DNS analyses, including a low-enthalpy, calorically perfect gas, and high-enthalpy multi-species gas mixtures under both chemical non-equilibrium and thermochemical non-equilibrium conditions.

The remainder of this paper is organized as follows. Section § 2 introduces the governing thermodynamic and chemical models, along with the numerical methods employed in this study. Section § 3 presents the wall flow quantities and the velocity statistics of the turbulent boundary layers as a validation of the numerical methods. Section § 4 investigates the mean temperature profiles and evaluates the validity of corresponding generalized Reynolds analogies under thermochemical non-equilibrium state. In Section § 5, the impact of thermochemical non-equilibrium on turbulence-chemistry interactions is analyzed, and the coupling between the turbulent fluctuations and vibrational energy relaxation is discussed. Section § 6 summarizes the primary conclusions of this work.

## 2. Physical model and numerical methods

### 2.1. *Governing equations and multi-species transport models*

The governing equations for high-enthalpy turbulent boundary layer flows incorporating thermochemical non-equilibrium effects are cast as follows,

$$\frac{\partial \rho_n}{\partial t} + \frac{\partial \rho_n u_k}{\partial x_k} = -\frac{\partial \mathcal{J}_{n,k}}{\partial x_k} + \dot{\omega}_{ch,n} \quad (n = 1, \ldots, N), \tag{2.1}$$

$$\frac{\partial \rho u_i}{\partial t} + \frac{\partial \rho u_i u_k}{\partial x_k} = -\frac{\partial p}{\partial x_i} + \frac{\partial \tau_{ik}}{\partial x_k}, \tag{2.2}$$

$$\frac{\partial \rho E}{\partial t} + \frac{\partial \rho H u_k}{\partial x_k} = \frac{\partial u_i \tau_{ik}}{\partial x_k} - \frac{\partial q_{tr,k}}{\partial x_k} - \frac{\partial q_{vib,k}}{\partial x_k} - \frac{\partial}{\partial x_k}\left(\sum_{n=1}^{N} \mathcal{J}_{n,k} h_n\right), \tag{2.3}$$

$$\frac{\partial \rho e_{vib}}{\partial t} + \frac{\partial \rho e_{vib} u_k}{\partial x_k} = -\frac{\partial q_{vib,k}}{\partial x_k} - \frac{\partial}{\partial x_k}\left(\sum_{n=1}^{N} \mathcal{J}_{n,k} e_{vib,n}\right) + \dot{\omega}_{vib}, \tag{2.4}$$

In the governing equations, $u_i$ represents the velocity component in the $x_i$ direction, where the indices $i = 1, 2, 3$ correspond to the streamwise ($x$), wall-normal ($y$), and spanwise ($z$) directions, respectively, with associated velocity components denoted by $u$, $v$, and $w$. The variables $\rho$, $E$, and $H$ denote the mixture density, specific total energy, and specific total enthalpy, respectively, while $\rho_n$ represents the partial density of the $n$-th species within a mixture containing $N$ total species. The species mass fraction $Y_n$ and mole fraction $X_n$ are defined as:

$$Y_n = \frac{\rho_n}{\rho}, X_n = \frac{\rho_n / W_n}{\sum_{n=1}^{N} (\rho_n / W_n)}. \tag{2.5}$$

with $W_n$ being the species molar mass. The thermodynamic pressure is determined Dalton's law of partial pressures,

$$p = \rho T R_u \sum_{n=1}^{N} \frac{Y_n}{W_n} = T \sum_{n=1}^{N} \rho_n R_n \tag{2.6}$$

where $R_n$ is the gas constant of the $n$-th species and $R_u = 8.314 \mathrm{J/(mol \cdot K)}$ is the universal gas constant. The total energy $E$ and the mixture vibrational energy $e_{vib}$ are calculated as:

$$E = H - p/\rho = \sum_{n=1}^{N} Y_n h_n + \frac{1}{2} u_i u_i - p/\rho, \quad e_{vib} = \sum_{n=1}^{N} Y_n e_{vib,n} \tag{2.7}$$

The thermodynamic properties of the high-temperature air species are computed by accounting for the distinct contributions of the translational-rotational and vibrational energy modes according to the models of Park (1989). Within this framework, the specific enthalpy and vibrational energy of the $n$-th species are evaluated as follows:

$$h_n\left(T, T_{vib}\right) = h_{n,ref} + \int_{T_{ref}}^{T} c_{p,n}^{tr}(T')\mathrm{d}T' + e_{vib,n}(T_{vib}), \tag{2.8}$$

$$e_{vib,n}\left(T_{vib}\right) = \begin{cases} h_n^{\mathrm{eq}}\left(T_{vib}\right) - \int_{T_{ref}}^{T_{vib}} c_{p,n}^{tr}(T')\mathrm{d}T' - h_{n,ref}, & \text{for diatomic molecules} \\ 0, & \text{for monatomic molecules.} \end{cases} \tag{2.9}$$

where $h_{n,ref}$ denotes the enthalpy of formation of the $n$-th species at the reference temperature $T_{ref} = 298.15\,\mathrm{K}$. The translational-rotational component of the specific heat at constant pressure, $c_{p,n}^{\mathrm{tr}}$, is defined as:

$$c_{p,n}^{tr} = \begin{cases} 7R_n/2, & \text{for diatomic molecules} \\ 5R_n/2, & \text{for monatomic molecules.} \end{cases} \tag{2.10}$$

Under thermal equilibrium conditions ($T = T_{vib}$), the equilibrium enthalpy $h_n^{\mathrm{eq}}(T)$ of the $n$-th species is computed using the curve-fitting polynomials of McBride *et al.* (2002):

$$\frac{h_n^{\mathrm{eq}}\left(T\right)}{R_n T} = -a_1 T^{-2} + a_2 \frac{\ln T}{T} + a_3 + a_4 \frac{T}{2} + a_5 \frac{T^2}{3} + a_6 \frac{T^3}{4} + a_7 \frac{T^4}{5} + \frac{b_1}{T}, \tag{2.11}$$

The strain rate tensor $S_{ij}$ and the viscous stress tensor $\tau_{ij}$ are expressed as

$$S_{ij} = \frac{1}{2}\left(\frac{\partial u_i}{\partial x_j} + \frac{\partial u_j}{\partial x_i}\right), \quad \tau_{ij} = 2\mu\left(S_{ij} - \frac{1}{3}\delta_{ij} S_{kk}\right), \tag{2.12}$$

where $\mu$ represents the dynamic viscosity of the gas mixture. The translational-rotational and vibrational heat flux, $q_i^{tr}$ and $q_i^{vib}$, obey the Fourier's law

$$q_i^{tr} = -\kappa \frac{\partial T}{\partial x_i}, \quad q_i^{vib} = -\kappa_{vib} \frac{\partial T_{vib}}{\partial x_i}. \tag{2.13}$$

with $\kappa$ and $\kappa_{vib}$ denoting the thermal conductivity and vibrational thermal conductivities of the gas mixture, respectively.

The species mass diffusion flux $\mathcal{J}_{n,i}$ is modeled by Fick's law of diffusion:

$$\mathcal{J}_{n,i} = -\rho D_n \frac{\partial Y_n}{\partial x_i} \tag{2.14}$$

where the effective multi-component diffusion coefficient $D_n$ is defined as

$$D_n = \frac{1 - X_n}{\sum_{s=1, s\neq n}^{N} \frac{X_s + \varepsilon}{D_{n,s}}}, \tag{2.15}$$

with $\varepsilon = 10^{-12}$ being a small regularization parameter preventing singularity. The dynamic viscosity $\mu$, translational-rotational thermal conductivity $\kappa$, and vibrational thermal conductivity $\kappa_{vib}$ of the gas mixture are evaluated using Wilke's semi-empirical mixing rules, which are standard in high-temperature gas dynamics and reacting flow simulations (Wilke 1950; Bird *et al.* 2002). The transport coefficients for each individual species (namely, the species viscosity $\mu_n$, thermal conductivities $\kappa_n, \kappa_{\mathrm{vib},n}$, and binary diffusion coefficients $D_{n,s}$) are computed according to the studies of Gupta *et al.* (1990) and Passiatore *et al.* (2022).

## 2.2. *Thermochemical models*

High-temperature air is modeled as a five-species ($N = 5$) mixture consisting of $N_2$, $O_2$, NO, N, and O, subject to five reversible chemical reactions ($M = 5$), following the finite-rate kinetic mechanisms (Park 1993; Park *et al.* 2001):

$$\begin{aligned}
&\mathrm{R1{:}\ O_2 + M \rightleftharpoons 2\,O + M},\\
&\mathrm{R2{:}\ N_2 + M \rightleftharpoons 2\,N + M},\\
&\mathrm{R3{:}\ NO + M \rightleftharpoons N + O + M},\\
&\mathrm{R4{:}\ NO + O \rightleftharpoons N + O_2},\\
&\mathrm{R5{:}\ O + N_2 \rightleftharpoons N + NO}\cdot
\end{aligned} \tag{2.16}$$

where M represents an arbitrary collision partner (third body) from the five gas-phase species. Specifically, reactions R1–R3 are the molecular dissociation and atomic recombination processes, while reactions R4 and R5 represent the exchange reactions of the extended Zeldovich mechanism governing nitric oxide (NO) formation. The net mass production rate of the $n$-th species, $\dot{\omega}_{ch,n}$, is determined by the law of mass action:

$$\dot{\omega}_{ch,n} = W_n \sum_{r=1}^{M} \left(\nu''_{nr} - \nu'_{nr}\right) \times \left[k_{f,r} \prod_{n=1}^{N} \left(\frac{\rho_n}{W_n}\right)^{\nu'_{nr}} - k_{b,r} \prod_{n=1}^{N} \left(\frac{\rho_n}{W_n}\right)^{\nu''_{nr}}\right] \tag{2.17}$$

where $\nu'_{nr}$ and $\nu''_{nr}$ are the stoichiometric coefficients of the $n$-th species acting as a reactant and a product in the $r$-th reaction step, respectively. The parameters $k_{f,r}$ and $k_{b,r}$ denote the forward and backward reaction rate coefficients of the $r$-th reaction, respectively, which are evaluated via the Arrhenius law. To account for the coupling between thermal and chemical non-equilibrium (vibrational-dissociation coupling), the forward rate coefficients of the dissociation reactions (R1-R3) are evaluated using Park's two-temperature model(Park 1993) at the controlling temperature $T_\mathrm{a} = \sqrt{TT_\mathrm{vib}}$. For the exchange reactions (R4 and R5), the controlling temperature is set to the translational-rotational temperature ($T_\mathrm{a} = T$).

The source term for the mixture vibrational energy, $\dot{\omega}_{vib}$, accounts for both the energy transfer between the translational-rotational and vibrational modes (translation-vibration relaxation) and the change in vibrational energy due to the production or depletion of molecular species by chemical reactions:

$$\dot{\omega}_{vib} = \dot{Q}_{vib} + \dot{Q}_{ch} = \dot{Q}_{vib} + \sum_{n=1}^{N} \dot{\omega}_n e_{vib,n} \tag{2.18}$$

The translation-vibration energy exchange rate is modeled as (Park *et al.* 2001)

$$\dot{Q}_{vib} = \sum_{m=1}^{N} \rho_m \frac{e_{vib,m}(T) - e_{vib,m}(T_{vib})}{\tau_m^{vib}}, \tag{2.19}$$

where the summation is restricted to the molecular (diatomic) species, as monatomic species lack vibrational degrees of freedom. The effective vibrational relaxation time of species $m$ in the multi-component mixture, $\tau_m^{vib}$, is computed as follows

$$\frac{1}{\tau_m^{vib}} = \sum_{n=1}^{N} \frac{X_n}{\tau_{mn}^{vib}} \tag{2.20}$$

The inter-species binary relaxation time $\tau_{mn}^{\mathrm{vib}}$ for the collision pair $m$-$n$ is calculated using the semi-empirical Millikan–White formulation, augmented with Park's high-temperature correction (Millikan & White 1963; Park 1993; Passiatore *et al.* 2022).

### 2.3. *Computational configuration and numerical methods*

The computational configuration for the high-enthalpy flow condition is schematically illustrated in Fig. 1. We consider a turbulent boundary layer developing over a 9° half-angle wedge under flight conditions corresponding to a freestream Mach number of 20 at an altitude of 36km. The flow parameters downstream of the leading-edge oblique shock wave are used as the boundary layer freestream conditions (hereinafter denoted by the subscript $\infty$). The post-shock freestream Mach number is $Ma_\infty = 9.74$, the unit Reynolds number is $Re_\infty = 4.9 \times 10^6 \mathrm{m}^{-1}$, and the static temperature is $T_\infty = 957.9$K, which is sufficiently high to initiate high-temperature gas effects. This post-shock freestream flow is assumed to be in thermochemical equilibrium and chemically frozen, yielding a freestream vibrational temperature of $T_{vib,\infty} = T_\infty$ and a composition of nitrogen and oxygen with mass fractions of $Y_{\mathrm{N_2},\infty} = 0.767$ and $Y_{\mathrm{O_2},\infty} = 0.233$, respectively. Under these conditions, the specific heat ratio of the gas mixture is $\gamma_\infty = 1.34$.

To isolate the distinct high-temperature phenomena, three simulation cases are investigated. Case HE-TC (High-Enthalpy Thermochemical Non-Equilibrium) accounts for both thermal and chemical non-equilibrium processes using the two-temperature model. Case HE-C (High-Enthalpy Chemical Non-Equilibrium) considers chemical non-equilibrium effects only, assuming local thermal equilibrium ($T = T_{vib}$). Case LE (Low-Enthalpy) is set as a comparative low-enthalpy case characterized by a significantly lower freestream temperature ($T_\infty = 287.7$K) but with comparable aerodynamic parameters. In this case, high-temperature gas effects are neglected, and the turbulent flow is governed by the classical Navier-Stokes equations for a calorically perfect gas with the dynamic viscosity $\mu$ determined via the Sutherland's law.

The computational domain is a rectangular cuboid configured with the following boundary conditions. At the flow inlet, a synthetic turbulence generator is implemented, superimposing resolved turbulent fluctuations onto mean profiles extracted from a precursor Reynolds-Averaged Navier-Stokes (RANS) simulation of the boundary layer under identical freestream conditions. The velocity and thermodynamic fluctuations are generated using the synthetic digital filtering method of Adler *et al.* (2018). The nominal boundary layer thickness at the inlet is $\delta_{\mathrm{in}} = 5$mm. At the outlet and top boundaries, non-reflecting characteristic boundary conditions are prescribed to prevent spurious pressure wave reflections. Periodic boundary conditions are imposed in the spanwise direction, assuming statistical homogeneity in the spanwise direction. At the wall, no-slip and no-penetration conditions are prescribed for the velocity, and the isothermal condition is enforced for the temperatures ($T = T_{vib} = T_w$), under a fully non-catalytic wall assumption for the species mass fractions, $\partial Y_n/\partial y|_w = 0$. Under these configurations, the wall-to-recovery temperature ratio is fixed at $T_w/T_r = 0.135$ across all simulations. This correspond to an isothermal wall temperature of $T_w = 2000$K for cases HE-TC and HE-C, and $T_w = 700$K for case LE.

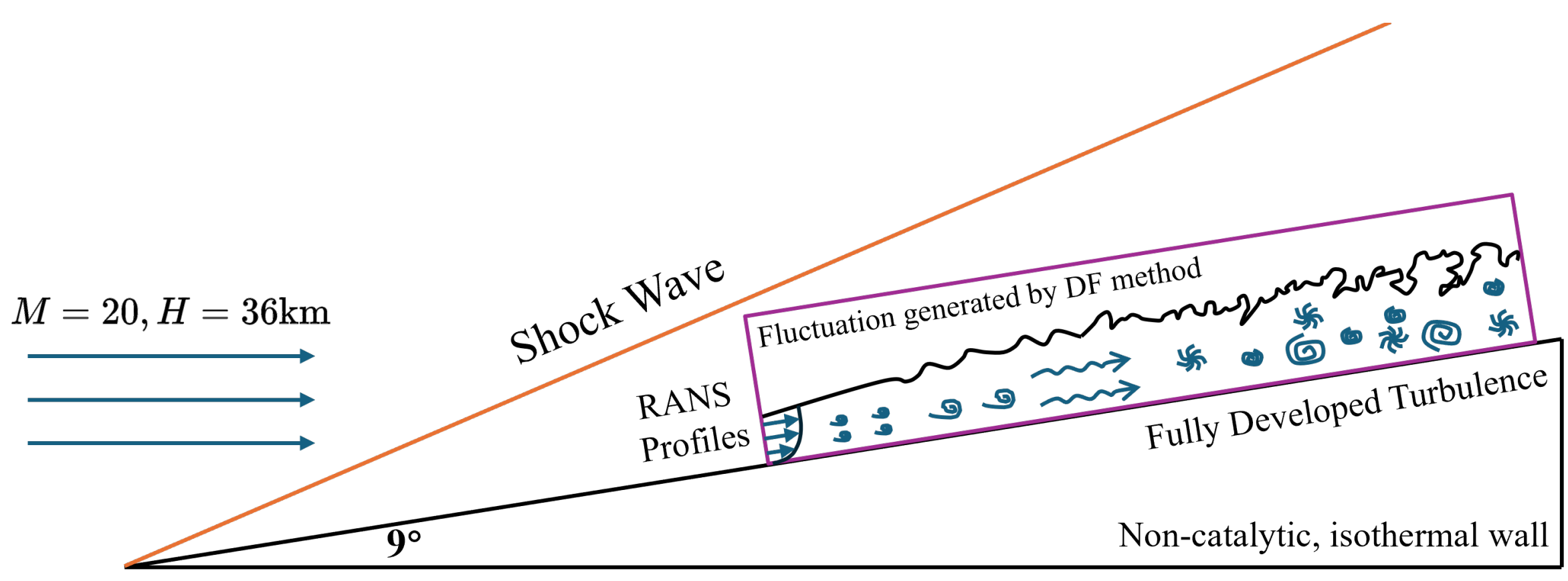

Figure 1: Sketch of high-enthalpy turbulent boundary layers.

The physical dimensions of the computational domain in three directions are given as $L_x \times L_y \times L_z = 204\delta_{in} \times 12\delta_{in} \times 10\delta_{in}$, which is discretized using a structured mesh containing $N_x \times N_y \times N_z = 3000 \times 360 \times 512$ grid points. To optimize computational efficiency while ensuring high-fidelity resolution in the region of interest, the streamwise grid distribution is divided into three distinct zones: the inlet development zone ($L_{x1} = 120\delta_{in}$), discretized by 750 uniformly distributed grid points, designed to allow the synthetic turbulence to spatially develop and transition into a statistically equilibrium turbulent state; a transition zone ($L_{x2} = 20\delta_{in}$), discretized by 250 grid points with a smoothly varying spacing, serving as a transitional buffer from the coarser upstream grid to the finer downstream grid; and the analysis zone ($L_{x3} = 64\delta_{in}$), discretized by 2000 uniformly distributed grid points, dedicated to capturing the fine-scale structures of the fully developed turbulent boundary layer. In the wall-normal direction, the grid points are clustered near the wall using an exponential stretching function with a constant stretching ratio of 1.0125, ensuring sufficient resolution of the viscous sublayer and near-wall gradients. Above the boundary layer, the grid spacing smoothly transitions to a uniform distribution in the freestream. In the spanwise direction, a uniform grid spacing is adopted.

Other notations used in the present study are introduced as follows. For a generic flow quantity $\phi$, the ensembled average (obtained by integrating over the statistically homogeneous spanwise direction and time) is denoted by $\bar{\phi}$, with the corresponding fluctuating component defined as $\phi'$. The density-avaraged (Favre) average is denoted by $\tilde{\phi}$, and the fluctuations by $\phi''$. Mean quantities evaluated at the wall are marked with the subscript $w$. The classical wall-shear scales (viscous units) are defined using the mean wall shear stress $\bar{\tau}_w = \mu_w(\partial\tilde{u}/\partial y)_w$, the mean wall density $\bar{\rho}_w$ and the mean wall dynamic viscosity $\bar{\mu}_w$. According to these flow quantities, the friction velocity $u_\tau$, the viscous length scale $\delta_\nu$, and the friction Reynolds

| Case | $Re_{\delta^*}$ | $Re_\theta$ | $Re_\tau$ | $10Ma_\tau$ | $H$ | $\Delta x^+$ | $\Delta z^+$ | $\Delta y_w^+$ | $\Delta y_\delta^+$ |
|---|---|---|---|---|---|---|---|---|---|
| HE-TC | 9637 | 2776 | 768 | 2.13 | 3.47 | 7.96 | 4.89 | 0.54 | 10.0 |
| HE-C | 9960 | 2913 | 766 | 2.14 | 3.42 | 7.94 | 4.87 | 0.54 | 9.99 |
| LE | 9861 | 2584 | 621 | 1.99 | 3.82 | 6.19 | 3.80 | 0.42 | 8.08 |

Table 1: Flow parameters and statistics at $x = 0.8$m. Here, $Re_{\delta^*}$ and $Re_\theta$ are the Reynolds number defined based on the displacement and momentum thicknesses, respectively. $Ma_\tau = u_\tau/\bar{c}_w$ is the friction Mach number. $H$ is the shape factor of the boundary layer.

number $Re_\tau$ are defined as

$$u_\tau = \sqrt{\frac{\bar{\tau}_w}{\bar{\rho}_w}}, \quad \delta_\nu = \frac{\bar{\mu}_w}{\bar{\rho}_w u_\tau}, \quad Re_\tau = \bar{\rho}_w u_\tau \delta/\bar{\mu}_w \tag{2.21}$$

with $\delta$ being the nominal boundary layer thickness. Variables normalized by these wall-viscous scales are indicated by the superscript +. To account for spatial variation of the mean density and viscosity across the boundary layer, the semi-local scaling variables (marked by the superscript $\star$) are employed by replacing their local mean values $\bar{\rho}$ and $\bar{\mu}$ by their wall counterparts,

$$u_\tau^\star = u_\tau \sqrt{\frac{\bar{\rho}_w}{\bar{\rho}}}, \quad \delta_\nu^\star = \frac{\bar{\mu}}{\bar{\rho} u_\tau^\star}. \tag{2.22}$$

Variables non-dimensionalized by these semi-local scales are marked by a superscript $\star$.

The DNS are performed using an in-house high-order finite-difference solver designed for reacting compressible turbulent flows. This solver has been extensively validated for high-speed wall-bounded turbulent flows in our previous studies (Li *et al.* 2022; Huang *et al.* 2025, 2026). The inviscid terms are discretized using a hybrid scheme that adaptively switches based on local flow conditions. In smooth turbulent regions, the seventh-order upwind difference scheme is used to minimize numerical dissipation, while a shock-capturing seventh-order weighted essentially non-oscillatory (WENO) scheme (Jiang & Shu 1996; Su *et al.* 2023) is activated near sharp flow discontinuities and shock waves. The viscous terms are computed using a central eighth-order finite-difference scheme. Time advancement is achieved with the explicit third-order total variation diminishing (TVD) Runge-Kutta scheme.

Table 1 summarizes the boundary layer parameters and turbulent statistics evaluated at the streamwise station $x = 0.8$ m, where the boundary layer is fully developed. At this location, the displacement Reynolds number $Re_{\delta^*}$ and the momentum Reynolds number $Re_\theta$ are approximately the same across all three cases. In contrast, the friction Reynolds number $Re_\tau$ is comparable between the high-enthalpy cases (HE-TC and HE-C) but is significantly lower for the low-enthalpy case LE. This deviation is primarily caused by the differences in the mean wall dynamic viscosity $\bar{\mu}_w$. The spatial resolutions evaluated in wall units are reported at this streamwise station. The uniform grid spacings in the streamwise and spanwise directions in the analysis zone are $\Delta x^+ \lesssim 8$ and $\Delta z^+ \lesssim 5$, respectively. In the wall-normal direction, the first grid point off the wall is positioned at the heights of $\Delta y_w^+ \lesssim 0.6$, and the grid spacing at the boundary layer edge is $\Delta y_\delta^+ \lesssim 10$. This grid resolution meets the requirement for direct numerical simulations, consistent with the established guidelines in the literature (Pirozzoli & Bernardini 2011; Zhang *et al.* 2018; Huang *et al.* 2020; Passiatore *et al.* 2022).

## 3. Basic flow statistics

In this section, we analyze the scalings of the basic boundary layer statistics, including the skin-friction coefficient, wall heat flux, mean velocity and Reynolds stresses. The analysis focuses on identifying the physical deviations introduced by high-enthalpy effects. Specifically, we assess the necessity of resolving vibrational non-equilibrium (via the two-temperature model) compared to assuming local thermal equilibrium, thereby isolating the influences of thermochemical non-equilibrium states on the turbulent structure.

### 3.1. *Skin friction and wall heat flux*

We begin by examining the skin-friction coefficient $C_f$ and the wall heat flux coefficient $C_h$. The skin-friction coefficient is defined as:

$$C_f = \frac{2\bar{\tau}_w}{\rho_\infty U_\infty^2} \tag{3.1}$$

and the wall heat flux coefficient is formulated as:

$$C_h = \frac{\bar{q}_{t,w}}{\rho_\infty U_\infty^3}, \quad \bar{q}_{t,w} = \bar{\kappa}_w \left.\frac{\partial \tilde{T}}{\partial y}\right|_w + \bar{\kappa}_{vib,w} \left.\frac{\partial \tilde{T}_{vib}}{\partial y}\right|_w + \sum_{n=1}^{N} \bar{\rho}_w \bar{D}_{n,w} \tilde{h}_{n,w} \left.\frac{\partial \tilde{Y}_n}{\partial y}\right|_w . \tag{3.2}$$

where the total mean wall heat flux in such thermochemically reacting flows consists of translational-rotational conduction, vibrational conduction, and species mass diffusion. Here, $\bar{\kappa}_{tr,w}$ and $\bar{\kappa}_{vib,w}$ represent the translational-rotational and vibrational thermal conductivities of the mixture at the wall, respectively. The $\bar{D}_{n,\mathrm{w}}$ is the effective diffusion coefficient of species $n$, and $\tilde{h}_{n,\mathrm{w}}$ is the enthalpy of species $n$. Under the fully non-catalytic wall conditions implemented in this study, the species mass fraction gradients at the wall vanish. Consequently, the species mass diffusion term in the heat flux equation is identically zero. The total wall heat flux can thus be decomposed into contributions from translational-rotational conduction and vibrational conduction $C_h = C_h^{tr} + C_h^{vib}$. Note that for case HE-C, the vibrational temperature equals the translational-rotational temperature ($T_{vib} = T$), whereas for case LE, the vibrational energy mode is unexcited, rendering the vibrational heat flux contribution to be zero.

Fig. 2 shows the streamwise distributions of the skin-friction coefficient $C_f$ and wall heat flux coefficient $C_h$ along the streamwise direction. The upstream region from the inlet to $x = 40\delta_{in}$ is omitted from the plots, as the turbulence in this zone is still in its spatial development state. The transitional development of the synthetic turbulence persists until approximately $x \approx 120\delta_{in}$, where the grid transitions to a finer resolution capable of resolving fine-scale near-wall structures. Downstream of $x \approx 120\delta_{in}$, both $C_f$ and $C_h$ flatten out and display a subtle, monotonic decrease with $x$ due to the spatial growth of the boundary layer thickness. Such extended streamwise recovery distances toward a fully developed, equilibrium turbulent boundary layer are typical when using digital-filtering synthetic turbulence generators, particularly for high-Mach-number flows with cold-wall boundary conditions. Comparing the different cases in Fig. 2(a,b), both high-enthalpy cases (HE-TC and HE-C) exhibit consistently higher $C_f$ and $C_h$ values than the low-enthalpy case (LE). Furthermore, the chemical non-equilibrium case under local thermal equilibrium (HE-C) yields slightly higher wall shear and heat transfer than the full thermochemical non-equilibrium case (HE-TC). These disparities stem primarily from the variations in near-wall fluid properties, specifically, density and dynamic viscosity, induced by high-enthalpy thermodynamics.

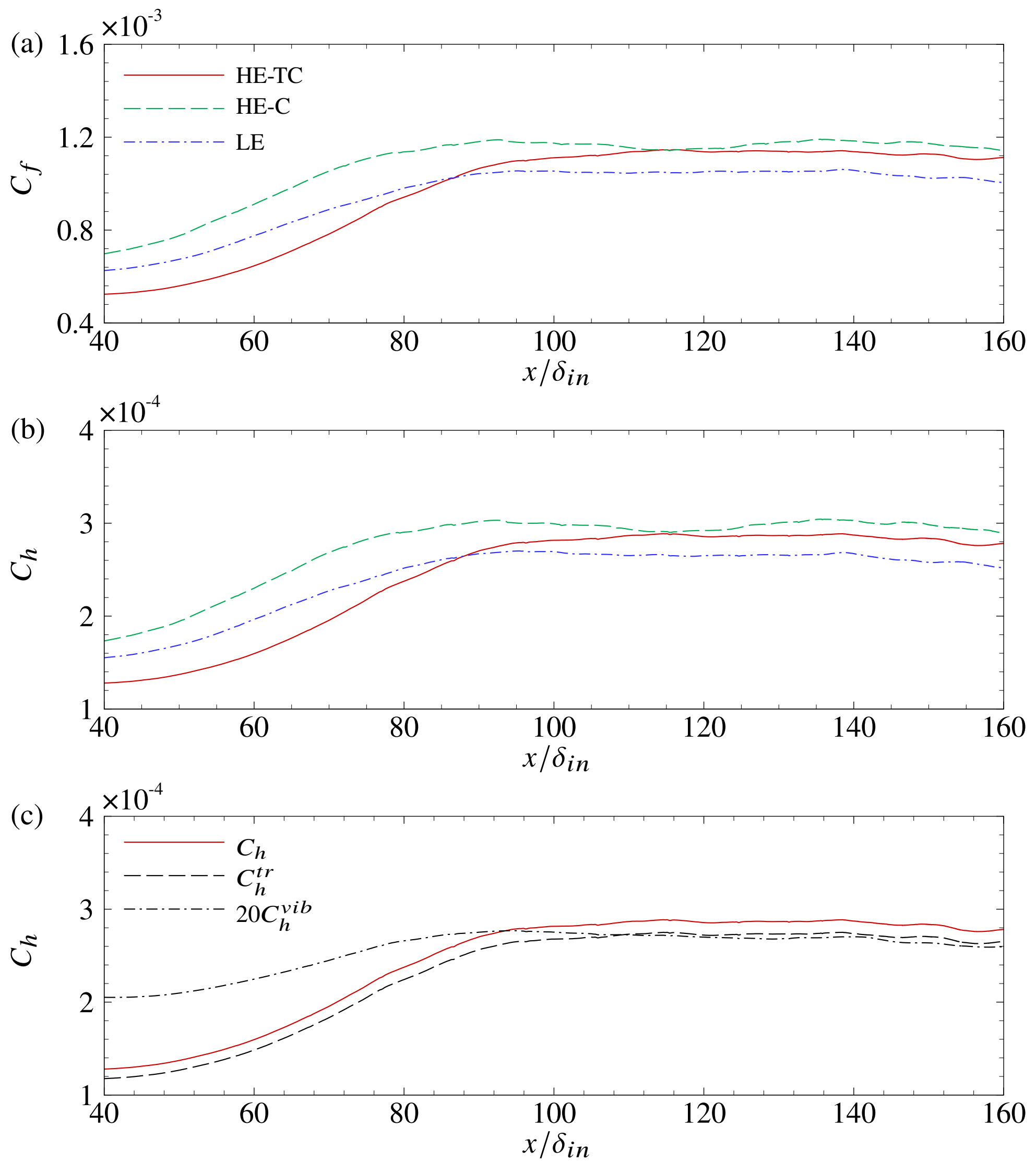

Figure 2: Streamwise evolution of (a) $C_f$ and (b) $C_h$ in all cases, and (c) $C_h$, $C_h^{tr}$ and $C_h^{vib}$ in case HE-TC.

For case HE-TC, the total heat flux coefficient $C_h$ is split into its translational-rotational component $C_h^{tr}$ and vibrational component $C_h^{vib}$. Because the mixture's vibrational thermal conductivity $\bar{\kappa}_{vib,w}$ is much smaller than its translational-rotational counterpart $\bar{\kappa}_{tr,w}$, the vibrational contribution $C_h^{vib}$ accounts for only approximately 5% of $C_h^{tr}$ in the fully developed region. Despite its small direct magnitude, thermal non-equilibrium effects cannot be ignored, for the temperature relaxation alters the forward and backward chemical reaction rates within the governing species transport equations. This alters the species concentration profiles across the boundary layer, which feedback-couples to the near-wall temperature gradients and ultimately the wall heat flux. This physical coupling mechanism will be detailed in subsequent discussions.

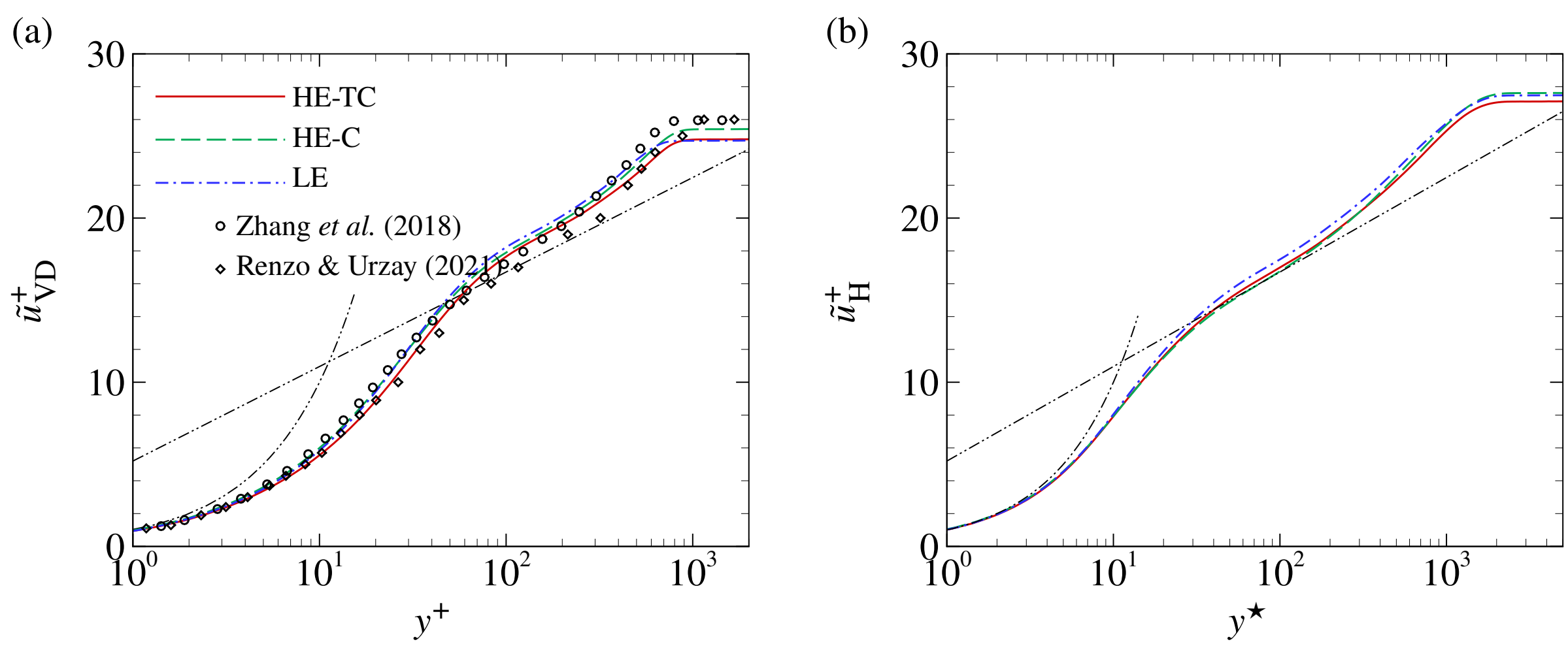

Figure 3: Wall-normal distributions of mean streamwise velocity under integral transformation (a) Equation (3.3), (b) Equation (3.4).

### 3.2. *Mean velocity and Reynolds stresses*

In high-Mach-number turbulent boundary layers, strong compressibility effects, manifested through significant mean density and viscosity gradients, alter the mean velocity profiles, causing them to deviate from the canonical incompressible law-of-the-wall. To account for these thermodynamic variations and recover a universal velocity scaling, integral transformations are traditionally applied. Fig. 3(a) presents the transformed mean velocity profiles using the classical van Driest (VD) transformation (van Driest 1951):

$$\tilde{u}^+_{\mathrm{VD}} = \int_0^{\tilde{u}^+} \sqrt{\frac{\bar{\rho}}{\rho_w}} \mathrm{d}\tilde{u}^+ \tag{3.3}$$

As established in prior studies (Pirozzoli & Bernardini 2011; Duan *et al.* 2011; Pirozzoli & Bernardini 2013), while the van Driest transformation succeeds in collapsing compressible profiles under quasi-adiabatic conditions, it fails to achieve Mach-number independence in strongly non-adiabatic, cold-wall configurations such as those investigated here. For direct comparison, reference DNS data from Zhang *et al.* (2018) for a perfect-gas flow ($M_\infty = 13.64$, $T_{\mathrm{w}}/T_{\mathrm{r}} = 0.18$, $Re_\tau \approx 646$) and from Renzo & Urzay (2021) for a high-enthalpy flow under chemical non-equilibrium ($M_\infty = 10$, $T_{\mathrm{w}}/T_{\mathrm{r}} = 0.086$, $Re_\tau \approx 961$) are also plotted in Fig. 3(a). Our DNS results show good agreement with these reference data sets in the viscous sublayer, buffer layer and logarithmic region. The minor discrepancies observed in the wake region ($y^+ > 100$) are primarily attributable to differences in the flow parameters, such as the freestream Mach number, wall temperature ratio $T_w/T_r$, and local friction Reynolds number $Re_\tau$.

To account for non-zero wall heat flux, several improved integral transformations have been proposed (Trettel & Larsson 2016; Volpiani *et al.* 2020; Griffin *et al.* 2021). Here, we evaluate the velocity scaling proposed by Hasan *et al.* (2023):

$$\tilde{u}_{\mathrm{H}}^{+}(y^{\star}) = \int_{0}^{\tilde{u}^{+}} \left(\frac{1 + k y^{\star} D^{c}}{1 + k y^{\star} D^{i}}\right) \left(1 - \frac{y}{\delta_{v}^{\star}} \frac{\mathrm{d}\delta_{v}^{\star}}{\mathrm{d}y}\right) \sqrt{\frac{\bar{\rho}}{\bar{\rho}_{w}}} \mathrm{d}\tilde{u}^{+} \tag{3.4}$$

where the van Driest damping functions for incompressible ($D^i$) and compressible ($D^c$) flows are expressed as

$$D^{i} = \left[1 - \exp\left(\frac{-y^{\star}}{A^{+}}\right)\right]^{2}, \; D^{c} = \left[1 - \exp\left(\frac{-y^{\star}}{A^{+} + f\left(M_{\tau}\right)}\right)\right]^{2} \tag{3.5}$$

with $A^+ = 17$ and $f(M_\tau) = 19.3M_\tau$. As shown in Fig. 3(b), applying the transformation of Hasan *et al.* (2023) successfully collapses the mean velocity profiles across all simulated flows onto the incompressible target curves, namely the linear law in the viscous sublayer and the logarithmic law in the log region. No discernible differences are observed among cases LE, HE-C, and HE-TC under this scaling. This collapse demonstrates that high-enthalpy physics, including both chemical non-equilibrium and thermochemical non-equilibrium effects, do not directly alter the mean momentum transport. Rather, high-enthalpy effects influence the velocity field indirectly via the spatial variation of mean thermodynamic properties ($\bar{\rho}$ and $\bar{\mu}$), adhering to the generalized compressibility scaling observed in cold-wall perfect-gas turbulent boundary layers (Duan *et al.* 2011; Duan & Martin 2011*b*; Zhang *et al.* 2018; Griffin *et al.* 2021).

Fig. 4 displays the wall-normal distributions of the root-mean-square (RMS) velocity fluctuations and the Reynolds shear stress, normalized using semi-local viscous scales. For comparative assessment, reference DNS datasets from Zhang *et al.* (2018) and Passiatore *et al.* (2022) ($M_\infty = 12.48$, $T_w/T_r = 0.10$, $Re_\tau \approx 1128$) are plotted alongside our results. The current cases (LE, HE-C, and HE-TC) are similar to the flow conditions of Zhang *et al.* (2018), exhibiting excellent agreement across the inner scaling region under semi-local viscous coordinates. Furthermore, when plotted against outer-scaled coordinates (omitted here for conciseness), the profiles yield excellent agreement in the outer region. This close match with high-fidelity reference databases validates both the grid resolution and the fidelity of the DNS databases established in this work.

## 4. Mean temperature and generalized Reynolds analogy

In this section, we examine the mean temperature field and its association with the mean velocity. Fig. 5(a) presents the wall-normal distributions of the mean translational-rotational temperature $\tilde{T}$, normalized by the freestream static temperature $T_\infty$, plotted against the semi-local coordinate $y^\star$. Under the prescribed cold-wall conditions, $\tilde{T}$ reaches a distinct local maximum within the buffer region at $y^\star \approx 10$ across all three cases. The peak values correspond to $4.6T_\infty$ for case HE-TC, $4.2T_\infty$ for case HE-C, and $5.3T_\infty$ for case LE, beyond which $\tilde{T}$ monotonically decays toward the freestream value.

Relative to the low-enthalpy reference case (LE), both high-enthalpy cases (HE-TC and HE-C) exhibit systematically lower non-dimensional mean temperatures throughout the inner and logarithmic regions ($y^\star < 400$, corresponding to $y/\delta < 0.56$). This temperature abatement is directly caused by the endothermic nature of molecular dissociation reactions occurring in the high-temperature near-wall region, which absorb sensible heat and convert thermal energy into chemical potential energy. In the outer boundary layer ($y^\star > 400$), high-enthalpy real-gas effects weaken, causing the normalized temperature profiles ($\tilde{T}/T_\infty$) of all three cases to collapse onto a single curve. Comparing the two high-enthalpy configurations, case HE-TC exhibits a higher mean translational-rotational temperature than case HE-C in the near-wall region ($y^\star < 150$, or $y/\delta < 0.3$). Physically, this occurs because thermal non-equilibrium causes the vibrational temperature to lag behind the translational-rotational temperature ($T_{\mathrm{vib}} < T$). This thermal lag lowers the effective controlling temperature governing forward dissociation reactions, thereby suppressing the rate of endothermic gas dissociation and leaving more thermal energy in the translational-rotational mode. Further away from the wall ($y^\star > 150$), local temperatures decrease, reaction rates become lower, and thermal non-equilibrium effects diminish, resulting in identical $\tilde{T}$ profiles between cases HE-TC and HE-C.

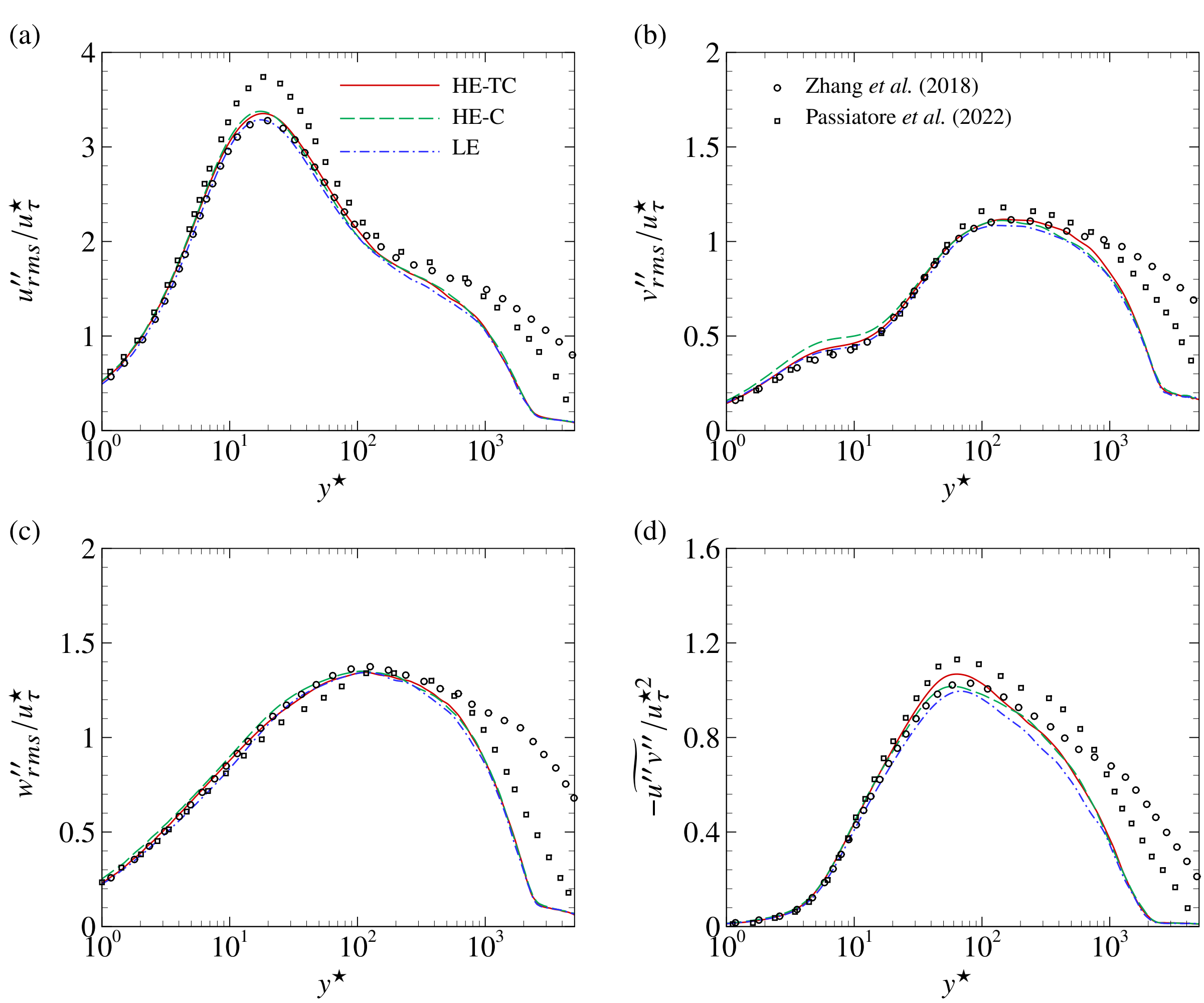

Figure 4: Wall-normal distributions of (a) $u''_{rms}$, (b) $v''_{rms}$, (c) $w''_{rms}$ and (d) $-\widetilde{u''v''}$ normalized by semi-local viscous scales.

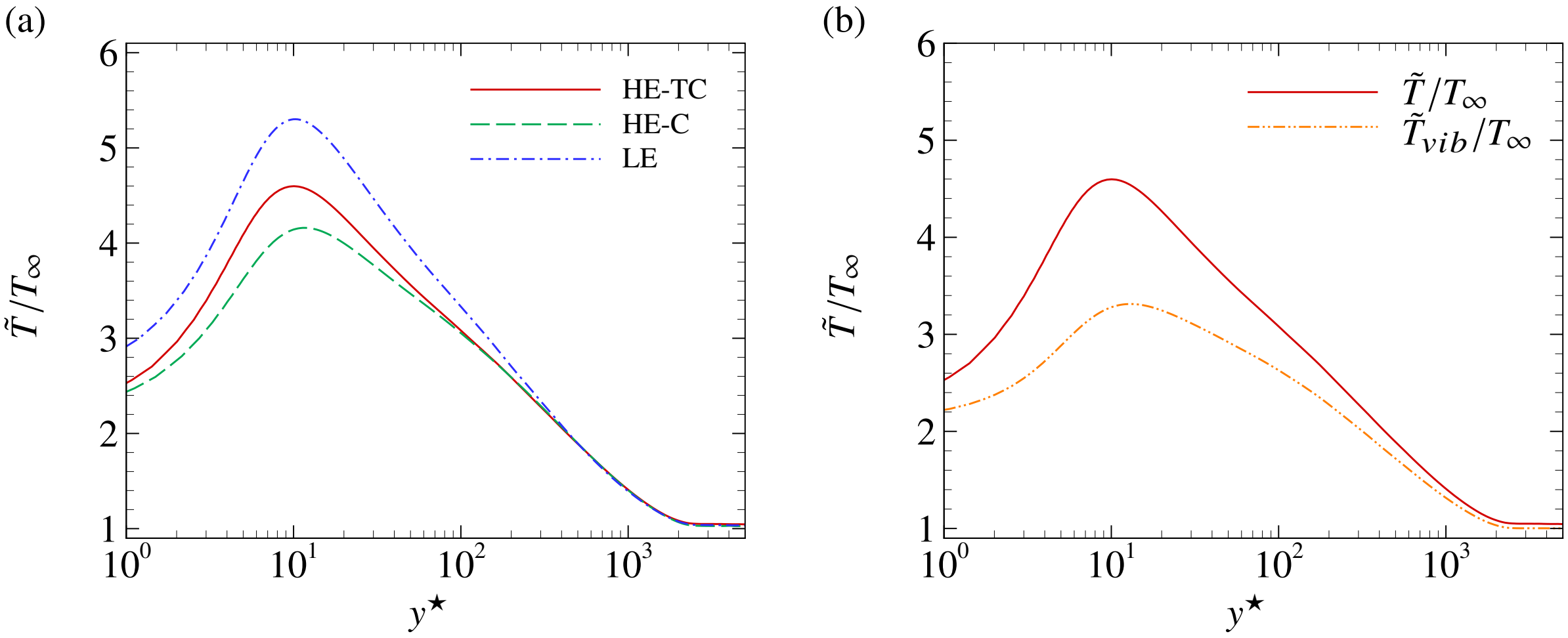

Figure 5: Wall-normal profiles of mean temperature, (a) $\tilde{T}/T_\infty$, (b) $\tilde{T}/T_\infty$ and $\tilde{T}_{vib}/T_\infty$ in case HE-TC.

Fig. 5(b) presents the wall-normal profile of the mean vibrational temperature $\tilde{T}_{vib}$ alongside the mean translational-rotational temperature $\tilde{T}$ for case HE-TC. Throughout the boundary layer, the translational-rotational temperature remains consistently higher than the vibrational temperature ($\tilde{T} > \tilde{T}_{vib}$), departing from thermal equilibrium except at the wall and in the freestream where thermal equilibrium is explicitly enforced. The thermal non-equilibrium degree, quantified by the temperature difference ($\tilde{T} - \tilde{T}_{vib}$), increases rapidly from the wall, reaching its maximum near the peak temperature location ($y^\star \approx 10$). At this peak, $\tilde{T}_{vib}$ reaches only approximately 70% of $\tilde{T}$ in absolute terms, highlighting strong local thermal non-equilibrium. Further away from the wall, this temperature difference gradually diminishes toward the boundary layer edge. This persistent departure between $\tilde{T}$ and $\tilde{T}_{\mathrm{vib}}$ alters the effective forward and backward reaction rates, directly suppressing chemical dissociation throughout the inner region of the boundary layer.

The relationship between the mean temperature and the mean velocity can be described using algebraic coupling relations. For a calorically perfect gas, invoking the balance of momentum and energy transfer under a constant Prandtl number leads to a quadratic relation between mean temperature and mean velocity (Busemann 1931; Crocco 1932; van Driest 1951; Walz 1962). A widely adopted formulation that accounts for non-unity Prandtl numbers and non-zero wall heat flux is the generalized Reynolds analogy (GRA) proposed by Zhang *et al.* (2014):

$$\frac{\tilde{T}}{\tilde{T}_\delta} = \frac{\tilde{T}_w}{\tilde{T}_\delta} + \frac{\tilde{T}_{rg} - \tilde{T}_w}{\tilde{T}_\delta}\frac{\tilde{u}}{\tilde{u}_\delta} + \frac{\tilde{T}_\delta - \tilde{T}_{rg}}{\tilde{T}_\delta}\left(\frac{\tilde{u}}{\tilde{u}_\delta}\right)^2, \tag{4.1}$$

where $\tilde{T}_{rg} = \tilde{T}_\delta + r_g\tilde{u}_\delta^2/(2c_p)$ represents the generalized recovery temperature, and $r_g$ is the generalized recovery factor. In high-enthalpy boundary layers, characterized by molecular excitation and dissociation, the specific heat capacity $c_p$ of the gas mixture varies spatially. Under these conditions, the GRA is more appropriately formulated by substituting static

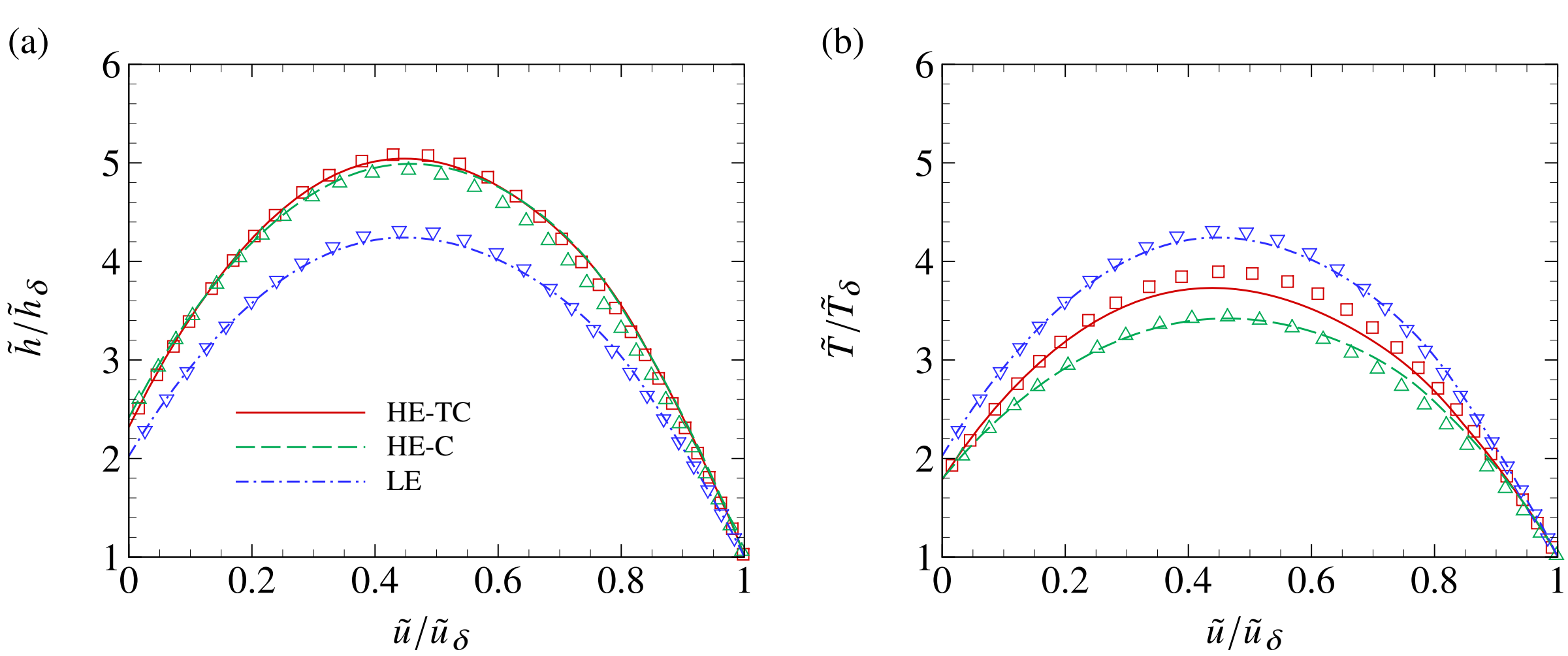

Figure 6: Distributions of (a) $\tilde{h}/\tilde{h}_\delta$ and (b) $\tilde{T}/\tilde{T}_\delta$ against $\tilde{u}/\tilde{u}_\delta$. Symbols: GRA relation.

temperature with mean static enthalpy $\tilde{h}$ (Duan & Martin 2011*b*; Li *et al.* 2022)

$$\frac{\tilde{h}}{\tilde{h}_\delta} = \frac{\tilde{h}_w}{\tilde{h}_\delta} + \frac{\tilde{h}_{rg} - \tilde{h}_w}{\tilde{h}_\delta}\frac{\tilde{u}}{\tilde{u}_\delta} + \frac{\tilde{h}_\delta - \tilde{h}_{rg}}{\tilde{h}_\delta}\left(\frac{\tilde{u}}{\tilde{u}_\delta}\right)^2, \tag{4.2}$$

where $\tilde{h}_{\mathrm{rg}}$ is the generalized recovery enthalpy, defined analogously to account for variable thermophysical properties.

In Fig. 6(a), we plot the mean static enthalpy profiles against the normalized mean velocity for all three cases, alongside the predictions from the enthalpy-based GRA relation (Equation 4.2). The enthalpy-based formulation exhibits excellent agreement with the DNS data across all cases, confirming its robustness in both low-enthalpy and high-enthalpy turbulent boundary layers. However, when the temperature-based GRA formulation (Equation 4.1) is evaluated, as shown in Figure 6(b), its validity varies between cases. While it remains highly accurate for cases LE and HE-C, a noticeable deviation from the DNS data occurs in case HE-TC within the range $0.3 < \tilde{u}/\tilde{u}_\delta < 0.7$, in agreement with the observations of Passiatore *et al.* (2022). This discrepancy stems directly from thermal non-equilibrium, in that the total static enthalpy incorporates a significant contribution from the vibrational energy, which is different from and cannot be characterized by the translational-rotational temperature $\tilde{T}$ alone.

Given that $\tilde{T}$ and $\tilde{T}_{vib}$ differ under thermochemical non-equilibrium conditions, it is of interest to examine whether the generalized Reynolds analogy can be adapted to predict the mean vibrational temperature profile $\tilde{T}_{vib}$. Fig. 7 plots the normalized mean vibrational temperature $\tilde{T}_{vib}/\tilde{T}_{vib,\delta}$ against the normalized mean velocity $\tilde{u}/\tilde{u}_\delta$, alongside the standard GRA formulation (Equation 4.1) directly evaluated using the vibrational boundary conditions. As shown, a direct extension of the standard GRA to the vibrational energy mode fails to capture the spatial variation of $\tilde{T}_{vib}$, with substantial deviations accumulating in the outer region of the boundary layer. This failure stems from the differing physical behaviors of the two temperature profiles across the boundary layer. In the inner region, where thermal non-equilibrium is most pronounced, the lag in vibrational excitation leads to a severe departure of $\tilde{T}_{vib}$ from $\tilde{T}$. Under these conditions, the departure cannot be adequately represented by a single quadratic of the mean velocity. Conversely, in the outer region, local temperatures decrease, causing reaction and relaxation times to lengthen relative to the flow timescales, yielding a gradual relaxation of the vibrational temperature toward the translational-rotational temperature, restoring local thermal equilibrium.

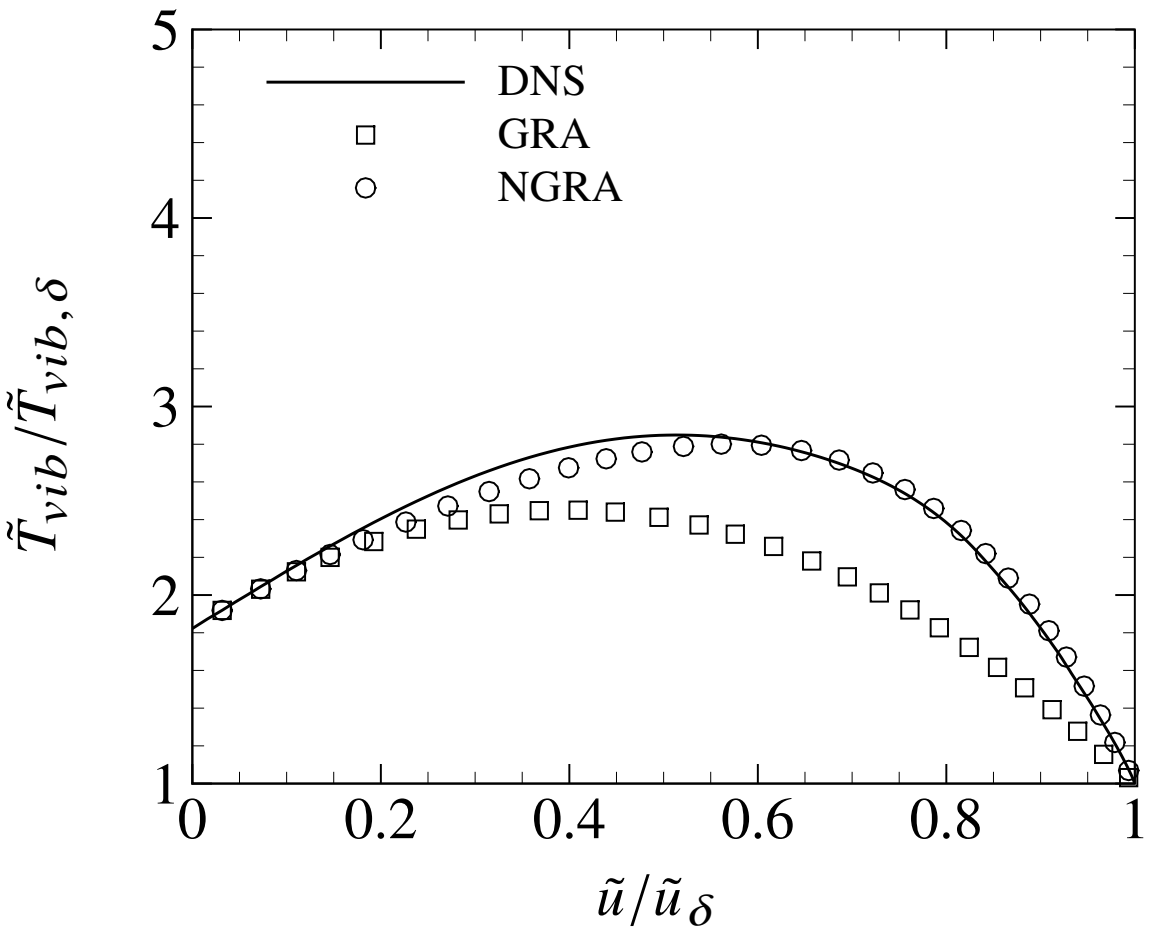

Figure 7: Wall-normal profiles of GRA in mean vibrational temperature in HE-TC case.

To model this behavior, we propose a composite formulation that blends the vibrational-temperature GRA with the standard translational-rotational temperature GRA:

$$\frac{\tilde{T}_{vib,\mathrm{NGRA}}}{\tilde{T}_{vib,\delta}} = \frac{\tilde{T}_{vib,\mathrm{GRA}}}{\tilde{T}_{vib,\delta}} \left[1 - \beta\left(\frac{\tilde{u}}{\tilde{u}_\delta}\right)\right] + \frac{\tilde{T}_{\mathrm{GRA}}}{\tilde{T}_\delta}\beta\left(\frac{\tilde{u}}{\tilde{u}_\delta}\right), \tag{4.3}$$

where $\tilde{T}_{vib,\mathrm{GRA}}$ represents the direct application of Equation 4.1 using vibrational temperature boundary conditions, and $\beta(\tilde{u}/\tilde{u}_\delta)$ is a blending weight function constrained by

$$\beta(0) = 0, \quad \beta(1) = 1, \quad \dot{\beta}(0) = 0. \tag{4.4}$$

These boundary conditions ensure that the composite model $\tilde{T}_{vib,\mathrm{NGRA}}$ matches the wall vibrational temperature and preserves the wall vibrational heat flux (since $\beta \to 0$ and its derivative vanishes at the wall), while asymptotically relaxing to the standard equilibrium GRA ($\tilde{T}_{\mathrm{GRA}}$) at the boundary layer edge ($\beta \to 1$). Approximating $\beta$ as a low-order polynomial satisfying these constraints yields the simple quadratic blending function leading to the final expression:

$$\frac{\tilde{T}_{vib,\mathrm{NGRA}}}{\tilde{T}_{vib,\delta}} = \frac{\tilde{T}_{vib,\mathrm{GRA}}}{\tilde{T}_{vib,\delta}} \left[1 - \left(\frac{\tilde{u}}{\tilde{u}_\delta}\right)^2\right] + \frac{\tilde{T}_{\mathrm{GRA}}}{\tilde{T}_\delta}\left(\frac{\tilde{u}}{\tilde{u}_\delta}\right)^2. \tag{4.5}$$

The predictions of this non-equilibrium GRA (NGRA) are evaluated in Fig. 7. The composite relation demonstrates excellent agreement with the DNS data across the entire boundary layer. This confirms that the proposed formulation successfully accounts for thermal non-equilibrium effects and accurately relates the mean vibrational temperature to the mean velocity field.

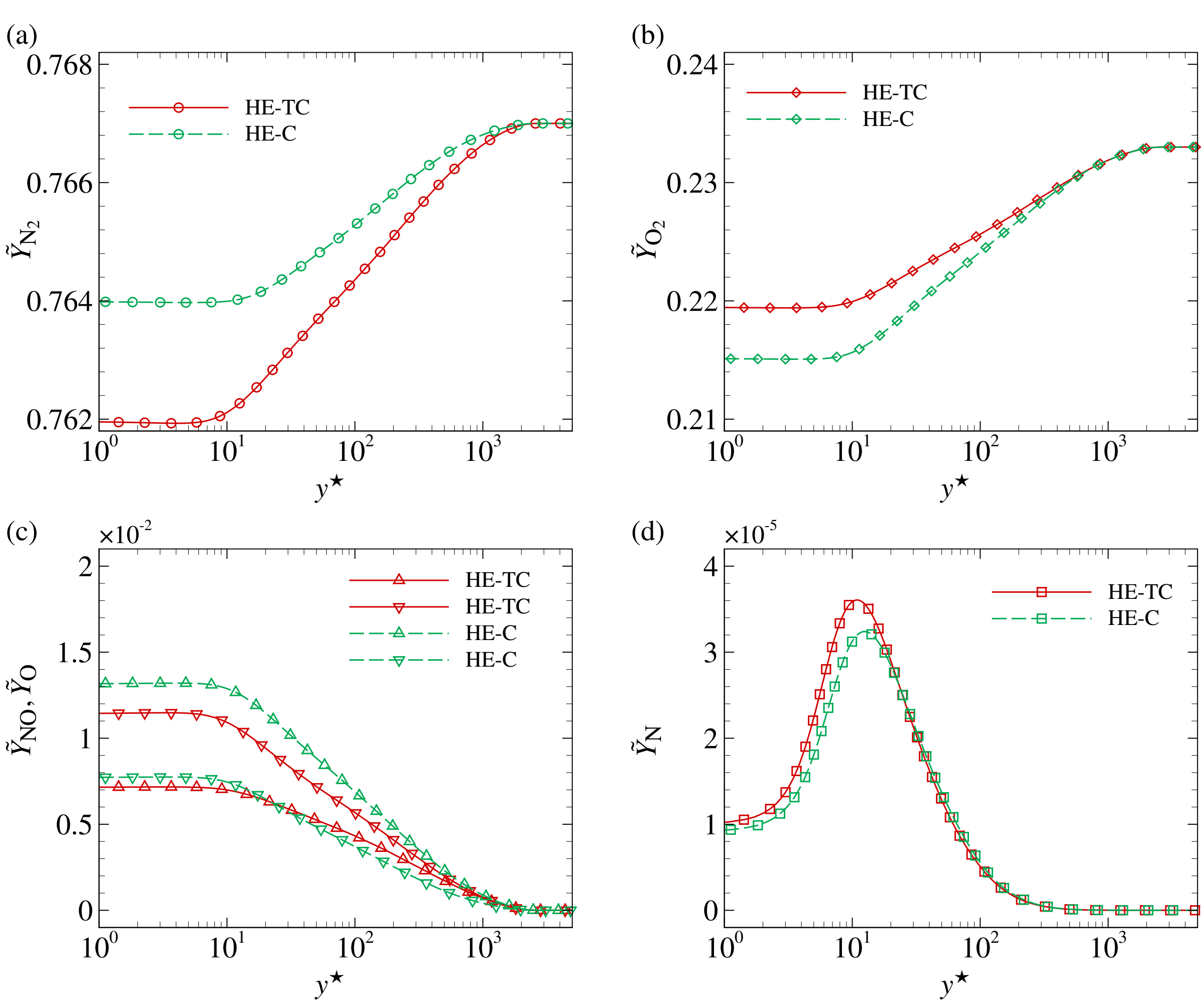

Figure 8: Wall-normal profiles of mean mass fraction, (a) $\tilde{Y}_{N_2}$, (b) $\tilde{Y}_{O_2}$, (c) $\tilde{Y}_{NO}$ (triangles) and $\tilde{Y}_O$ (downward triangles), (d) $\tilde{Y}_N$.

## 5. Thermochemical nonequilibrium effects on chemical reactions

In this section, we investigate the influence of thermochemical non-equilibrium on the mean mass fractions of the species within the gas mixture and examine the spatial behavior of the chemical source terms in the governing conservation equations, which dictate the localized rates of species production and destruction under high-enthalpy conditions.

Fig. 8 displays the wall-normal profiles of the mean mass fractions for the five chemical species. For the flow conditions under consideration, the peak mean temperature within the boundary layer is approximately 4500K, so the primary chemical processes are the dissociation of molecular oxygen and the subsequent production of atomic oxygen (O) via reaction R1, which reaches a peak mass fraction of approximately 0.7% (Figs. 8a–c). Nitric oxide (NO), synthesized via the Zel'dovich exchange mechanism (reactions R4–R5), constitutes approximately 1.1% of the local mixture mass. For both of these reaction products, the maximum mass fractions are attained at the wall.

A comparison between cases HE-C and HE-TC reveals that neglecting thermal non-equilibrium effects (case HE-C) leads to higher local mass fractions of O but lower mass fractions of NO. The difference in O production is directly attributable to thermal non-equilibrium effects. In case HE-TC, the lag in vibrational excitation reduces the controlling reaction temperature, conventionally modeled via the Park geometric mean $T_a = \sqrt{TT_{vib}}$, thereby suppressing the forward oxygen dissociation rate. Conversely, the Zel'dovich exchange reactions are primarily governed by the translational-rotational temperature $T$. Because the vibrational energy mode acts as an active energy sink under thermal equilibrium (case HE-C), the localized translational-rotational temperature $T$ is slightly lower compared to case HE-TC, which in turn leads to a minor reduction in NO production. Additionally, nitrogen dissociation is only beginning to initiate at these temperatures, rendering the mass fraction of atomic nitrogen (N) negligible (Fig. 8d), with virtually no discrepancy between cases HE-TC and HE-C.

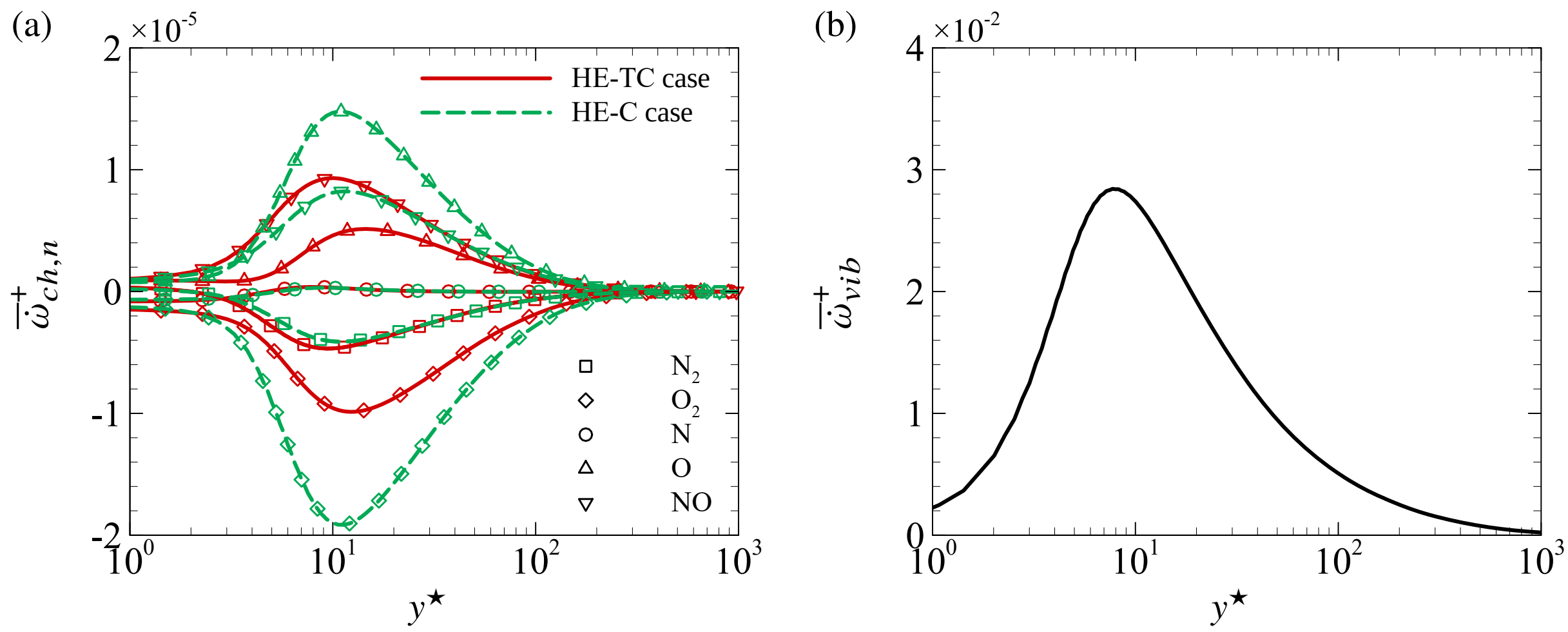

Figure 9: Wall-normal distributions of (a) $\overline{\dot{\omega}}^{+}_{ch,n}$ and (b) $\overline{\dot{\omega}}^{+}_{vib}$.

We also note that the fluctuations of translational-rotational temperature ($T''$), vibrational temperature ($T''_{vib}$), and species mass fractions ($Y''_n$) share highly similar structural features. These fluctuating fields act as passive scalars transported by energy-containing structures. This behavior is reflected in the turbulent Prandtl number and turbulent Schmidt numbers remaining close to unity across most of the boundary layer. Under these conditions, the strong Reynolds analogy, which correlates these scalar fluctuation intensities with the streamwise velocity fluctuations $u''$, remains qualitatively valid. Consequently, a detailed analysis of these fluctuating scalar fields is omitted here for brevity.

The spatial variations in the mean mass fractions of the constituent species are caused by the local chemical reaction rates. To elucidate this, Fig. 9 presents the wall-normal profiles of the normalized mean chemical source terms $\overline{\dot{\omega}}^{+}_{ch,n}$ for cases HE-TC and HE-C (Fig. 9a), alongside the mean vibrational energy source term $\overline{\dot{\omega}}^{+}_{vib}$ for case HE-TC (Fig. 9b). As shown in Fig. 9(a), the magnitudes of the chemical source terms peak in the near-wall region at $y^{\star} \approx 10$, which closely coincides with the location of the maximum mean translational-rotational temperature. Comparing the two cases, it is evident that incorporating thermal non-equilibrium (case HE-TC) only marginally enhances the net rate of conversion of molecular nitrogen $N_2$ into atomic nitrogen N and nitric oxide NO. In contrast, the magnitude of $\overline{\dot{\omega}}^{+}_{ch,n}$ for both $O_2$ and O is substantially higher under thermal equilibrium state (case HE-C). This pronounced difference confirms that thermal non-equilibrium strongly inhibits the oxygen dissociation process in the boundary layer, a direct consequence of the vibrational temperature lag which reduces the effective activation temperature of the forward reaction.

The mean vibrational energy source term $\overline{\dot{\omega}}_{vib}^{+}$, shown in Fig. 9(b), reaches its peak magnitude slightly closer to the wall at $y^{\star} \approx 8$. This location corresponds approximately to the region of maximum thermal non-equilibrium where the disparity between $\tilde{T}$ and $\tilde{T}_{vib}$ is the largest (as previously shown in Fig. 5b). The concentration of $\overline{\dot{\omega}}_{vib}^{+}$ in this high-shear layer underscores the intense translation-vibrational (T-V) energy exchange as the gas mixture relax toward local thermal equilibrium.

To gain deeper-level insights into these processes, the species chemical production rates $\dot{\omega}_{ch,n}$ can be split according to their underlying reaction pathways

$$\dot{\omega}_{ch,n} = \dot{\omega}_{ch,n}^{D}(\rho_s, T, T_{vib}) + \dot{\omega}_{ch,n}^{Z}(\rho_s, T) \tag{5.1}$$

where $\dot{\omega}_{ch,n}^{D}$ represents the contribution from the dissociation and recombination reactions (R1–R3), which depend on both the local species densities $\rho_s$ and the dual-temperature state $(T, T_{vib})$. The term $\dot{\omega}_{ch,n}^{Z}$ accounts for the contribution from the exchange reactions (R4–R5), which are assumed to be governed exclusively by the translational-rotational temperature $T$. In a similar manner, the total vibrational energy source term can be decomposed to isolate the individual physical mechanisms:

$$\dot{\omega}_{vib} = \dot{Q}_{vib} + \dot{Q}_{ch}, \tag{5.2}$$

where $\dot{Q}_{vib}$ represents the translation-vibrational (T-V) energy transfer rate, and $\dot{Q}_{ch}$ denotes the rate of change of vibrational energy directly caused by the creation or destruction of molecular species during chemical reactions.

The contributions of these decomposed terms to the species chemical production rates are shown in Figs.10(a,b). The chemical production rate of $O_2$ is jointly governed by both mechanisms, whereas that of atomic oxygen is dominated by the dissociation term $\dot{\omega}_{ch,\mathrm{O}}^{D}$. Conversely, the production and exchange of $N_2$, N, and NO are almost exclusively determined by the Zel'dovich exchange term $\dot{\omega}_{ch,n}^{Z}$. In the absence of thermal non-equilibrium effects (case HE-C), the forward rate of the dissociation reactions is overpredicted due to the overestimation of the controlling reaction temperature, resulting in substantially higher mean production rates for both $O_2$ and O.

Figs.10(c,d) display the decomposed contributions to the vibrational energy source terms. The total vibrational source term $\dot{\omega}_{vib}$ is determined almost entirely by the translation-vibrational energy transfer term $\dot{Q}_{vib}$, which reaches its maximum at $y^{\star} \approx 8$. In contrast, the chemical reaction term $\dot{Q}_{ch}$ is roughly two orders of magnitude smaller, indicating its secondary role under the presently considered flow conditions.

It is noteworthy that these local production rates are highly nonlinear functions of the temperatures and species mass fractions. This strong nonlinearity precludes their accurate estimation using simple mean flow quantities in Reynolds-Averaged Navier-Stokes (RANS) approaches, or filtered resolved-scale quantities in large-eddy simulations (LES). Consequently, the high-order correlations arising from subgrid-scale or unresolved fluctua-tions, commonly referred to as turbulence-chemistry interactions (TCI), must be rigorously characterized and modeled to achieve predictive accuracy in low-order simulations.

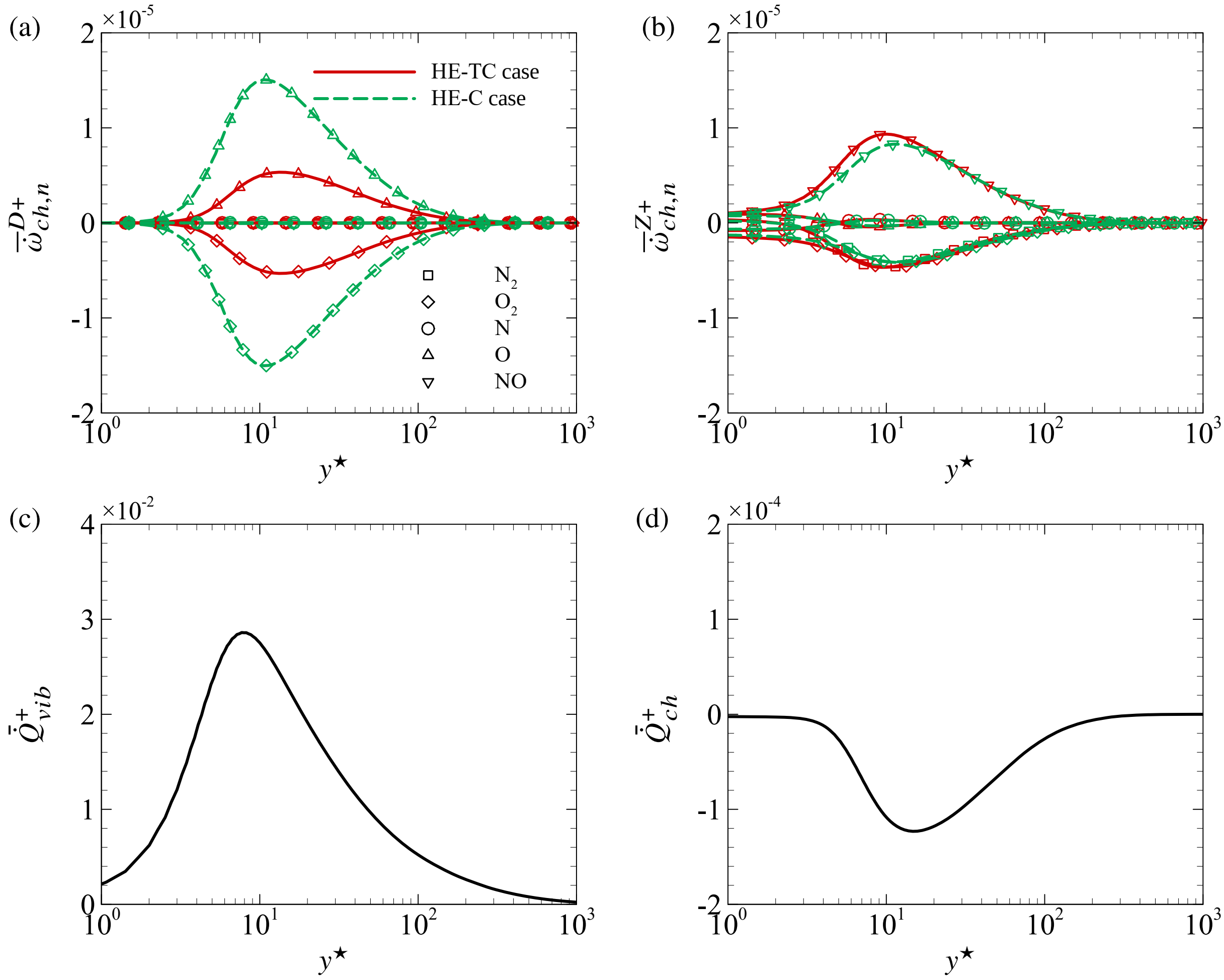

Figure 10: Wall-normal distributions of (a) $\overline{\dot{\omega}}_{ch,n}^{D+}$, (b) $\overline{\dot{\omega}}_{ch,n}^{Z+}$, (c) $\bar{\dot{Q}}_{vib}^{+}$ and (d) $\bar{\dot{Q}}_{ch}^{+}$

Within the context of RANS modeling, the unresolved contribution of turbulent fluctuations to the mean chemical production rates is isolated by defining the TCI term as

$$\dot{\omega}_{ch,n}^{I} = \overline{\dot{\omega}}_{ch,n} - \dot{\omega}_{ch,n}(\bar{\rho}_s, \tilde{T}, \tilde{T}_{vib}), \tag{5.3}$$

in which the second term on the right-hand side represents the 'laminar' rate evaluated directly using the mean density $\bar{\rho}_s$ and the Favre-averaged temperatures $\tilde{T}$ and $\tilde{T}_{vib}$. To isolate the individual and coupled roles of species density and temperature fluctuations, this interaction term is decomposed as follows (Duan & Martín 2011; Duan & Martin 2011*a*; Williams *et al.* 2025; Wang & Xu 2026)

$$\dot{\omega}_{ch,n}^{I} = \dot{\omega}_{ch,n}^{I,\rho} + \dot{\omega}_{ch,n}^{I,T} + \dot{\omega}_{ch,n}^{I,c}, \tag{5.4}$$

where the individual contributing components are formulated as

$$\dot{\omega}^{I,\rho}_{ch,n} = \overline{\dot{\omega}_{ch,n}(\rho_s, \tilde{T}, \tilde{T}_{vib})} - \dot{\omega}_{ch,n}(\bar{\rho}_s, \tilde{T}, \tilde{T}_{vib}), \tag{5.5a}$$

$$\dot{\omega}^{I,T}_{ch,n} = \overline{\dot{\omega}_{ch,n}(\bar{\rho}_s, T, T_{vib})} - \dot{\omega}_{ch,n}(\bar{\rho}_s, \tilde{T}, \tilde{T}_{vib}), \tag{5.5b}$$

$$\dot{\omega}^{I,c}_{ch,n} = \overline{\dot{\omega}_{ch,n}(\rho_s, T, T_{vib})} - \overline{\dot{\omega}_{ch,n}(\rho_s, \tilde{T}, \tilde{T}_{vib})} \tag{5.5c}$$

$$- \overline{\dot{\omega}_{ch,n}(\bar{\rho}_s, T, T_{vib})} + \dot{\omega}_{ch,n}(\bar{\rho}_s, \tilde{T}, \tilde{T}_{vib}), \tag{5.5d}$$

denoting the contribution arising solely from species-density fluctuations, temperature fluctuations (both $T$ and $T_{vib}$), the cross-correlation term capturing the coupled effects of density and temperature fluctuations, respectively.

Similarly, the turbulence-vibrational relaxation interaction (TVI) term is defined to quantify the influence of turbulent fluctuations on vibrational energy relaxation:

$$\dot{\omega}^{I}_{vib} = \overline{\dot{\omega}}_{vib} - \dot{\omega}_{vib}(\bar{\rho}_s, \tilde{T}, \tilde{T}_{vib}). \tag{5.6}$$

This term is similarly decomposed into constituent parts

$$\dot{\omega}^{I}_{vib} = \dot{\omega}^{I,\rho}_{vib} + \dot{\omega}^{I,T}_{vib} + \dot{\omega}^{I,c}_{vib}, \tag{5.7}$$

where the individual components are given as

$$\dot{\omega}^{I,\rho}_{vib} = \overline{\dot{\omega}_{vib}(\rho_s, \tilde{T}, \tilde{T}_{vib})} - \dot{\omega}_{vib}(\bar{\rho}_s, \tilde{T}, \tilde{T}_{vib}), \tag{5.8a}$$

$$\dot{\omega}^{I,T}_{vib} = \overline{\dot{\omega}_{vib}(\bar{\rho}_s, T, T_{vib})} - \dot{\omega}_{vib}(\bar{\rho}_s, \tilde{T}, \tilde{T}_{vib}), \tag{5.8b}$$

$$\dot{\omega}^{I,c}_{vib} = \overline{\dot{\omega}_{vib}(\rho_s, T, T_{vib})} - \overline{\dot{\omega}_{vib}(\rho_s, \tilde{T}, \tilde{T}_{vib})} \tag{5.8c}$$

$$- \overline{\dot{\omega}_{vib}(\bar{\rho}_s, T, T_{vib})} + \dot{\omega}_{vib}(\bar{\rho}_s, \tilde{T}, \tilde{T}_{vib}). \tag{5.8d}$$

The physical interpretations of these terms are the same as those for $\dot{\omega}^{I}_{ch,n}$.

Fig. 11 presents the wall-normal profiles of the decomposed TCI terms. Across all species, the temperature fluctuation term $\dot{\omega}^{I,T}_{ch,n}$ is clearly dominant. This finding is consistent with prior studies (Williams *et al.* 2025; Wang & Xu 2026), which concluded that the unresolved chemical source term is primarily governed by the strong nonlinear sensitivity of the Arrhenius reaction rates to temperature fluctuations. The cross-correlation term $\dot{\omega}^{I,c}_{ch,n}$ is also non-negligible, particularly for the key oxygen species ($O_2$ and O), although representing this term analytically poses a significant modeling challenge. In contrast, the contribution from species-density fluctuations $\dot{\omega}^{I,\rho}_{ch,n}$ is at least one order of magnitude smaller than that of the temperature fluctuations, suggesting it can be neglected under the present flow conditions.

When comparing the thermodynamic states, the TCI terms for case HE-TC exhibit not only lower overall magnitudes than those in case HE-C, but also distinct spatial distributions. This is evidenced by the noticeably different off-wall locations of their respective peak values. These differences indicate that thermal non-equilibrium effects play a critical role in modulating the coupling between turbulence and chemical reactions. Neglecting these non-equilibrium effects to simplify the computational model would lead to a substantial overestimation of the chemical production rates, even if the turbulent closures for the fluid dynamics were mathematically exact.

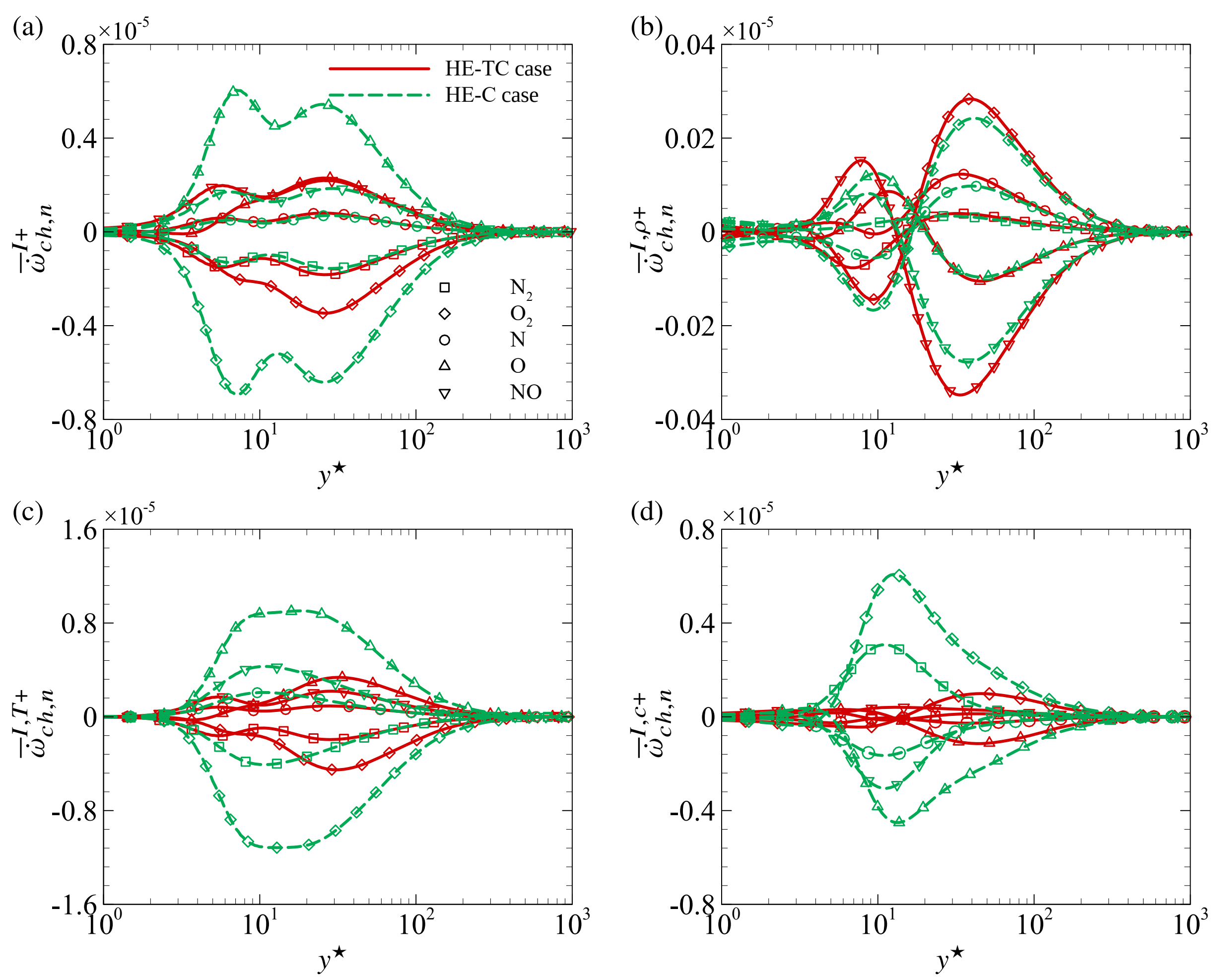

Figure 11: Wall-normal distributions of (a) $\overline{\dot{\omega}}^{I+}_{ch,n}$ (b) $\overline{\dot{\omega}}^{I,\rho+}_{ch,n}$, (c) $\overline{\dot{\omega}}^{I,T+}_{ch,n}$ and (d) $\overline{\dot{\omega}}^{I,c+}_{ch,n}$.

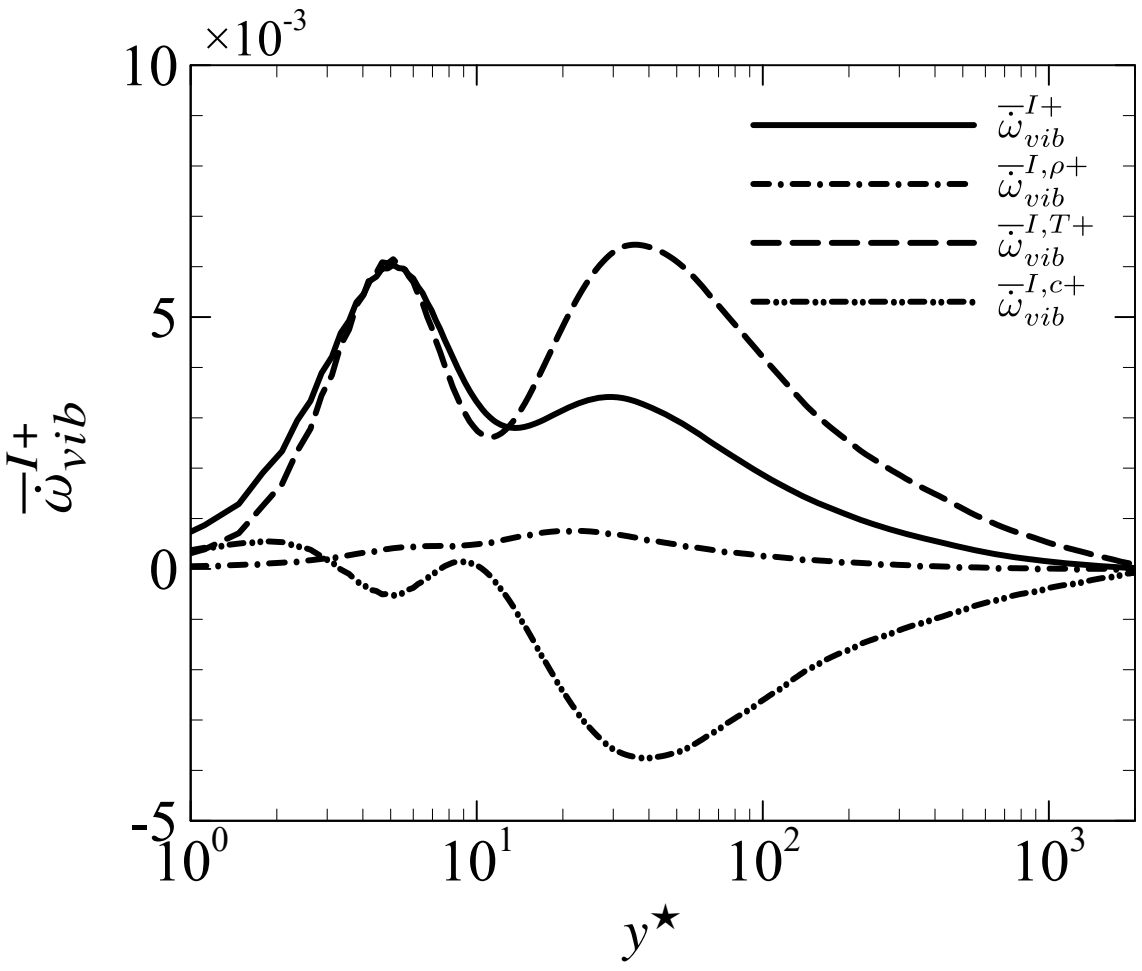

Figure 12: Wall-normal distributions of $\dot{\omega}^{I}_{vib}$ and its decomposition.

The decomposed components of the TVI term $\dot{\omega}^{I}_{vib}$ for case HE-TC are depicted in Fig. 12. In the inner region ($y^\star < 10$), the overall interaction term is dominated by the temperature fluctuation component $\dot{\omega}^{I,T}_{vib}$, with the species-density and cross-correlation terms remaining comparatively small. Further away from the wall ($y^\star > 10$), $\dot{\omega}^{I,T}_{vib}$ exceeds the magnitude of the total TVI term $\dot{\omega}^{I}_{vib}$, but this discrepancy is largely offset by a negative contribution from the cross-correlation term $\dot{\omega}_{vib}^{I,c}$. Similar to the behavior observed for the chemical production rates, the species-density fluctuation term $\dot{\omega}_{vib}^{I,\rho}$ remains negligible across the boundary layer.

In LES, subgrid-scale (SGS) closures are required to represent the unresolved contributions of the turbulence-chemistry and turbulence-vibrational relaxation interaction terms. To analyze how turbulent fluctuations at different spatial scales contribute to these interaction terms, we perform a scale-by-scale decomposition using a spectral-filtering approach in the homogeneous spanwise direction $z$. First, we apply a sharp low-pass filter in the spanwise wavenumber space with a variable cutoff wavenumber $k_c$ to reconstruct the resolved-scale fluctuations of the species densities, translational-rotational temperature, and vibrational temperature $T_c'' = \int_0^{k_c} \widehat{T''}\, \mathrm{d}k$, $T_{vib,c}'' = \int_0^{k_c} \widehat{T_{vib}''}\, \mathrm{d}k$ and $\rho_{s,c}'' = \int_0^{k_c} \widehat{\rho_s''}\, \mathrm{d}k$, where $\widehat{(\cdot)}$ denotes the spanwise Fourier transformation. Using these scale-by-scale fluctuations, we define the cumulative spectral contribution densities for the TCI and TVI terms by differentiating the filtered mean production rates with respect to the cutoff wavenumber

$$\widehat{\Phi}_{ch,n}^{I}(k_z) = \frac{\partial}{\partial k_z} \overline{\dot{\omega}_{ch,n}(\bar{\rho}_s + \rho_{s,c}', \tilde{T} + T_c'', \tilde{T}_{vib} + T_{vib,c}'')}, \tag{5.9}$$

$$\widehat{\Phi}_{vib}^{I}(k_z) = \frac{\partial}{\partial k_z} \overline{\dot{\omega}_{vib}(\bar{\rho}_s + \rho_{s,c}', \tilde{T} + T_c'', \tilde{T}_{vib} + T_{vib,c}'')}. \tag{5.10}$$

To isolating the primary influence of the temperature fluctuations on these scale-dependent terms, we define the corresponding temperature-fluctuation spectral densities as:

$$\widehat{\Phi}_{ch,n}^{I,T}(k_z) = \frac{\partial}{\partial k_z} \overline{\dot{\omega}_{ch,n}(\bar{\rho}_s, \tilde{T} + T_c'', \tilde{T}_{vib} + T_{vib,c}'')}, \tag{5.11}$$

$$\widehat{\Phi}_{vib}^{I,T}(k_z) = \frac{\partial}{\partial k_z} \overline{\dot{\omega}_{vib}(\bar{\rho}_s, \tilde{T} + T_c'', \tilde{T}_{vib} + T_{vib,c}'')}. \tag{5.12}$$

The spectral densities corresponding to the species-density fluctuations and the cross-correlation terms are formulated similarly. Integrating these spectral distribution functions over all wave numbers directly yields the total interaction terms.

Fig. 13 presents the pre-multiplied spanwise spectra of the turbulence-chemistry interaction terms for O and NO, which serve as the most representative species in the reacting mixture. Consistent with the integrated profiles discussed previously, the pre-multiplied spectra $k_z\widehat{\Phi}_{ch,\mathrm{NO}}^{I+}$ exhibits a distinct double-peak feature. The first peak is located very close to the wall at $y^\star \approx 5$, while the second is situated in the lower logarithmic region at $y^\star \approx 40$. In contrast, $k_z\widehat{\Phi}_{ch,\mathrm{O}}^{I+}$ displays a single peak. Notably, the peak contributions for both species align at a spanwise wavelength of $\lambda_z^\star \approx 200$. These spectral peaks correspond to the local maxima of the temperature fluctuations ($T''$), suggesting that temperature fluctuations at the energy-containing scales are an important component contributing the TCI. This link is further confirmed by the pre-multiplied spectra of the temperature-fluctuation component, $k_z\widehat{\Phi}_{ch,n}^{I,T+}$, which closely resemble the overall spatial and spectral distribution of $k_z\widehat{\Phi}_{ch,n}^{I+}$, albeit with larger magnitudes. This overestimation is counterbalanced by the negative contribution of the cross-correlation spectra $k_z\widehat{\Phi}_{ch,n}^{I,c+}$. Intriguingly, this cross-correlation term is highly active in the outer region for O, but is localized much closer to the wall for NO. This difference indicates that the production of these two species are modulated by distinct turbulent-chemistry coupling mechanisms at different distances from the wall.

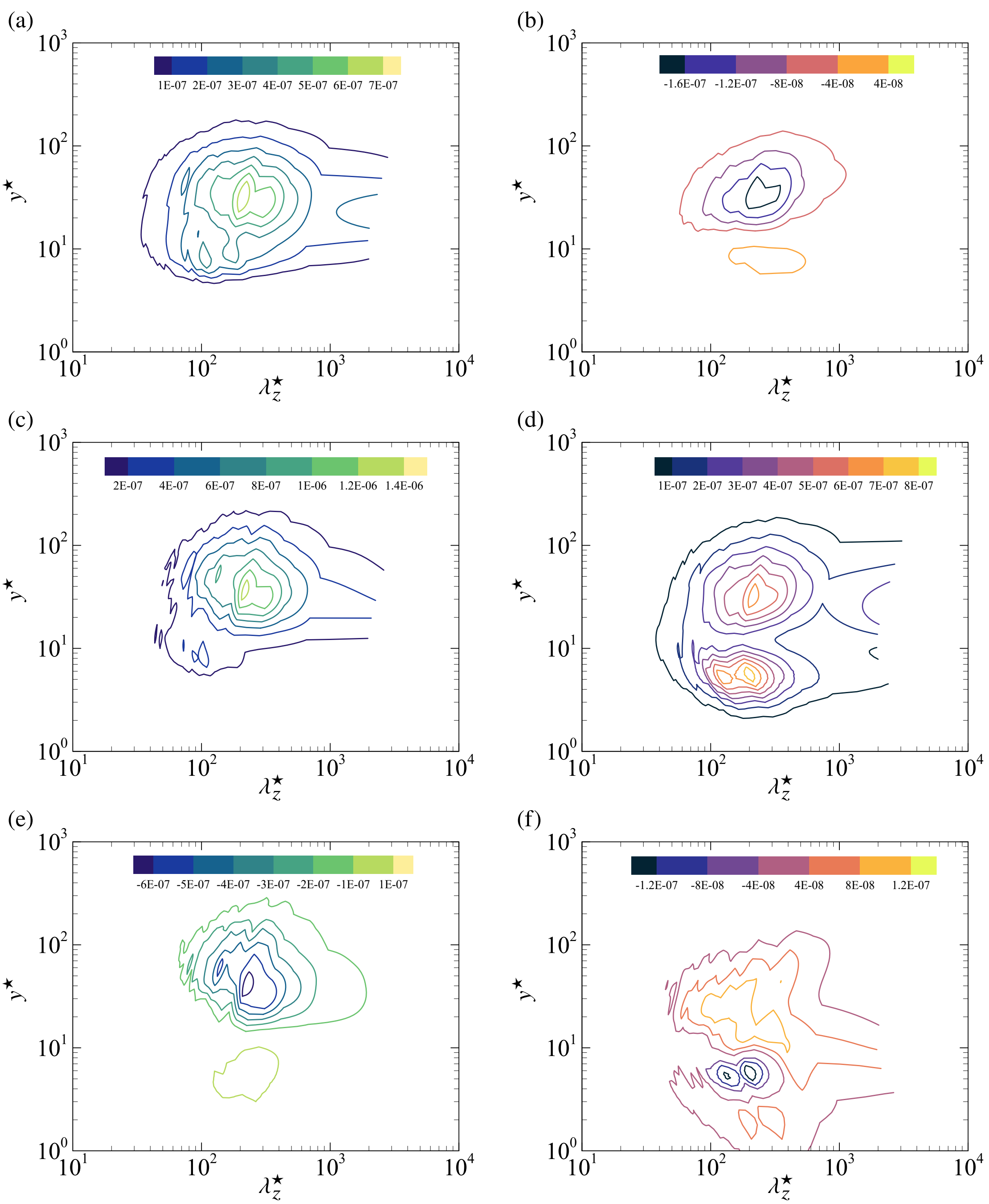

Figure 13: Pre-multiplied spectral functions of turbulence-chemistry interaction term for O and NO in case HE-TC, (a) $k_z\widehat{\Phi}^{I+}_{ch,\mathrm{O}}$, (b) $k_z\widehat{\Phi}^{I+}_{ch,\mathrm{NO}}$, (c) $k_z\widehat{\Phi}^{I,T+}_{ch,\mathrm{O}}$, (d) $k_z\widehat{\Phi}^{I,T+}_{ch,\mathrm{NO}}$, (e) $k_z\widehat{\Phi}^{I,c+}_{ch,\mathrm{O}}$, (f) $k_z\widehat{\Phi}^{I,c+}_{ch,\mathrm{NO}}$.

Integrating these spectral density functions from the highest resolved wavenumber down to a specified cutoff wavenumber $k_c$ isolates the small-scale contribution to the turbulence-chemistry interaction, hereafter denoted as the subgrid-scale (SGS) TCI term, $\tilde{\omega}^{I+}_{ch,n}$. This term represents the exact subgrid closure that must be modeled mathematically in LES.

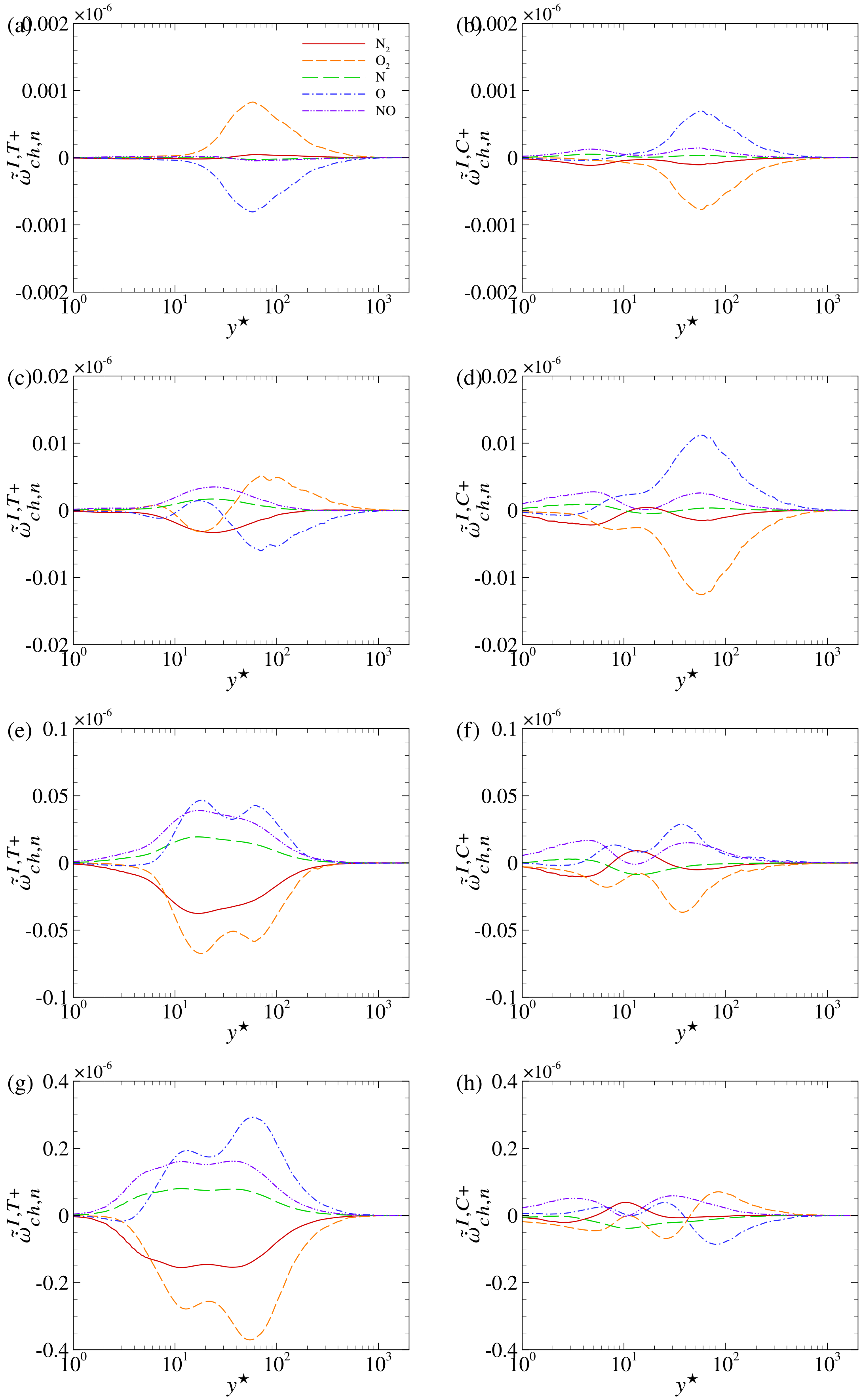

Figure 14: Wall-normal distributions of (a,c,e,g) $\tilde{\omega}^{I,T+}_{ch,n}$ (b,d,f,h) $\tilde{\omega}^{I,C+}_{ch,n}$ at the filtering scale of (a,b) $\lambda_z^+ = 25$, (c,d) $\lambda_z^+ = 50$, (e,f) $\lambda_z^+ = 100$, (g,h) $\lambda_z^+ = 200$.

Fig. 14 displays the subgrid temperature-fluctuation term $\tilde{\omega}^{I,T+}_{ch,n}$ and the subgrid cross-correlation term $\tilde{\omega}^{I,C+}_{ch,n}$, which constitute the two primary components of $\tilde{\omega}^{I+}_{ch,n}$, evaluated at spanwise cutoff wavelengths of $\lambda^+_z = 25$, 50, 100, and 200, respectively. At small filter widths ($\lambda^+_z = 25$ and 50), the cross-correlation term $\tilde{\omega}^{I,C+}_{ch,n}$ is of comparable, or even superior, magnitude to the temperature-fluctuation term $\tilde{\omega}^{I,T+}_{ch,n}$. This relative dominance suggests that directly applying simplified RANS-based subgrid models, such as the assumed probability density function (PDF) method (Wang & Xu 2026), would introduce significant inaccuracies because they typically neglect or simplify these cross-correlations. However, it is vital to recognize that at these fine filter widths ($\lambda^+_z \leqslant 50$), the absolute magnitude of the SGS TCI term is two to three orders of magnitude smaller than its scale-resolved counterpart. This indicates that while modeling the SGS TCI term at these scales is highly complex, it is practically unnecessary for the flow conditions investigated here. Conversely, at larger filter widths ($\lambda^+_z = 100$ and 200), the aggregate SGS TCI term becomes considerably more significant. In this coarse-filtering regime, the temperature-fluctuation term dominates the cross-correlation term, thereby restoring the validity and applicability of simplified PDF-based modeling approaches. These conditions are representative of wall-modeled LES, where the near-wall grid resolution is inherently coarser than in wall-resolved LES.

Fig. 15 presents the pre-multiplied spanwise spectra of the TVI term. Consistent with the integrated profiles discussed previously, $k_z\widehat{\Phi}^{I+}_{vib}$ displays a single prominent peak in the lower logarithmic region at $y^\star \approx 40$, occurring at a spanwise wavelength of $\lambda^\star_z \approx 200$. In terms of its decomposed components, the temperature-fluctuation term $k_z\widehat{\Phi}^{I,T+}_{vib}$ exhibits a double-peak feature with local maxima located near the wall at $y^\star \approx 5$, $\lambda^\star_z \approx 200$ and at $y^\star \approx 40$, $\lambda^\star_z \approx 200$, which correspond directly to the peak locations of the temperature fluctuations. The spectral magnitudes of $k_z\widehat{\Phi}^{I+}_{vib}$ are systematically larger than those of the total term $k_z\widehat{\Phi}^{I+}_{vib}$, with this excess being offset by the negative contribution of the cross-correlation spectra $k_z\widehat{\Phi}^{I,c+}_{vib}$. Meanwhile, the species-density fluctuation term $k_z\widehat{\Phi}^{I,\rho+}_{vib}$ remains at least two orders of magnitude smaller than the other two components across all scales. Overall, the spectral characteristics of the TVI term closely resemble those of the TCI term. Consequently, the previous analysis regarding the scale-dependent behavior of the subgrid-scale terms and the necessity of subgrid modeling at different filter widths applies equally to the TVI term.

## 6. Conclusions

In this study, direct numerical simulations (DNS) of high-Mach-number turbulent boundary layers were performed under three distinct thermochemical conditions: a low-enthalpy calorically perfect gas (LE), a high-enthalpy mixture in chemical non-equilibrium state (HE-C), and a high-enthalpy mixture in thermochemical non-equilibrium state (HE-TC). The influence of the two-temperature formulation on velocity and temperature statistics, as well as the turbulence-chemistry and turbulence-vibrational relaxation interactions, was systematically investigated. The main conclusions are summarized as follows.

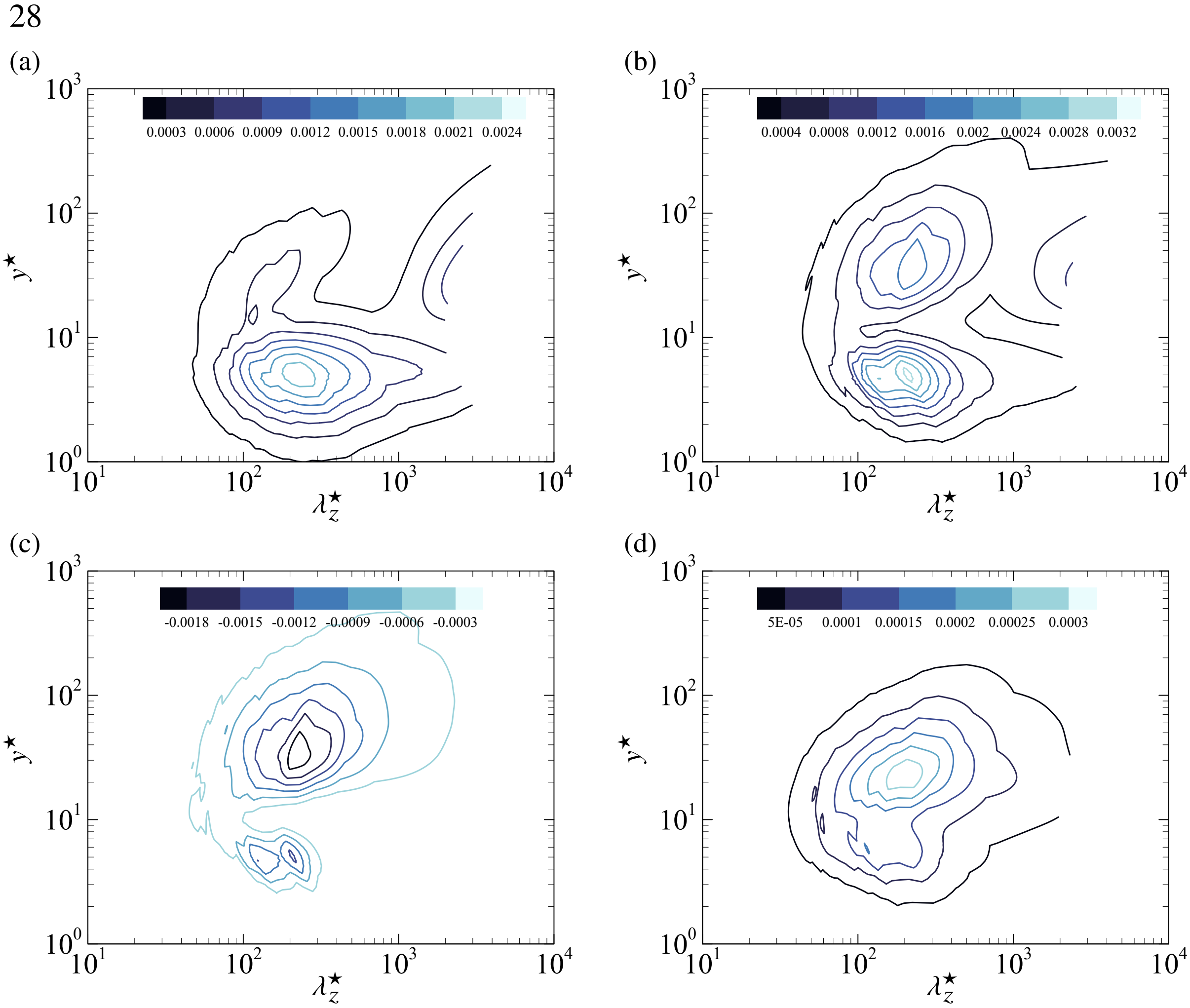

Figure 15: Pre-multiplied spectra of turbulence-vibrational relaxation interaction term in HE-TC case, (a) $k_z\widehat{\Phi}^{I+}_{vib}$, (b) $k_z\widehat{\Phi}^{I,T+}_{vib}$, (c) $k_z\widehat{\Phi}^{I,c+}_{vib}$ and (d) $k_z\widehat{\Phi}^{I,\rho+}_{vib}$.

High-enthalpy effects on statistics of boundary layers is firstly considered. High-enthalpy effects exert a minimal influence on the turbulent velocity statistics, including the mean velocity profiles and Reynolds stresses, but profoundly alter the near-wall thermal field. When normalized by the freestream temperature, the mean temperature is lower in the high-enthalpy cases than in the low-enthalpy reference case, with the HE-TC case exhibiting a slightly higher mean temperature than the HE-C counterpart. While the standard generalized Reynolds analogy (GRA) accurately describes the relationships between mean static enthalpy and velocity, as well as mean temperature and velocity, it fails to capture the vibrational temperature-velocity relation. Near the wall, strong thermal non-equilibrium effects induce a pronounced disparity between the Favre-averaged vibrational temperature $\tilde{T}_{vib}$ and the translational-rotational temperature $\tilde{T}$, whereas the two temperatures gradually equilibrate in the outer region. To address this, a composite GRA formulation is proposed by blending the vibrational-temperature and standard translational-rotational-temperature relations. Its excellent agreement with the DNS data demonstrates its capability to predict the mean vibrational-temperature-velocity relationship.

Modulations of near-wall chemical kinetics and energy relaxation are further discussed. Thermal non-equilibrium effects significantly modify the near-wall chemical kinetics. Among the primary product species, the formation of NO is dominated by exchange reactions, whereas O production is governed primarily by dissociation. By suppressing $O_2$ dissociation and promoting NO formation, thermal non-equilibrium reduces the mean O mass fraction and increases the mean NO mass fraction in the near-wall region. Consequently, both the mean mass production rate of O and its associated turbulence-chemistry interaction (TCI) term are suppressed, whereas the corresponding quantities for NO show a modest increase. Furthermore, the thermal non-equilibrium triggers vibrational energy relaxation, representing the energy exchange between the translational-rotational and vibrational modes, which leads to substantial vibrational energy source terms and strong turbulence-vibrational relaxation interactions (TVI).

Spectral characteristics and subgrid scale-by-scale interactions are explored in detail. Turbulent fluctuations contribute substantially to the mean chemical reaction and vibrational energy production rates. Decomposing the TCI and TVI terms into contributions from species-density fluctuations, temperature fluctuations, and their cross-correlations reveals that the temperature-fluctuation terms are dominant, while the cross-correlations remain non-negligible. A scale-by-scale spectral decomposition demonstrates that at energy-containing scales, temperature fluctuations dictate the behavior of both TCI and TVI. This analysis also confirms that O and NO chemical productions are determined by different mechanisms. Integrating these spectral distributions from the highest wavenumber down to a cutoff wavenumber yields the subgrid-scale (SGS) TCI and TVI terms, which represent the closures required for large-eddy simulations. At small filter widths (typical of wall-resolved LES), the magnitude of the subgrid cross-correlation term is comparable to, or even exceeds, that of the temperature-fluctuation term, though the absolute magnitude of the subgrid contribution is small compared to the resolved counterpart. Conversely, at larger filter widths, the subgrid contribution increases substantially, with the temperature-fluctuation term becoming the dominant component.

The scale-by-scale analysis underscores the necessity of developing scale-dependent SGS closure models for LES. Such models must represent the cross-correlation terms accurately, especially at small filter widths where standard models like the assumed PDF approach fail. This will be considered in our future work.